\documentclass{aa}  

\usepackage{caption}
\usepackage{color}
\usepackage{enumitem}
\usepackage{graphicx}
\usepackage{natbib}
\usepackage{comment}
\usepackage{txfonts}
\usepackage[colorlinks = true,
            linkcolor = blue,
            urlcolor  = blue,
            citecolor = blue,
            anchorcolor = blue]{hyperref}
\defcitealias{Giri2025}{Paper I}

\begin{document}

  \title{Probing the formation of megaparsec-scale giant radio galaxies\\ II. Continuum \& polarization behavior from MHD simulations}
  \titlerunning{Emission and Polarization signatures of GRGs}

   \author{Gourab Giri,\inst{1,2}
   Christian Fendt,\inst{3}
   Joydeep Bagchi,\inst{4}
   Kshitij Thorat,\inst{2}
   D. J. Saikia,\inst{5}
   Roger P. Deane,\inst{2,6}
   Jacinta Delhaize\inst{7}
          }

   \institute{
   Istituto Nazionale di Astrofisica (INAF) – Istituto di Radioastronomia (IRA), via Gobetti 101, 40129 Bologna, Italy
   \and
   Department of Physics, University of Pretoria, Private Bag X20, Hatfield 0028, South Africa
   \and
   Max Planck Institute for Astronomy, K\"onigstuhl 17, D-69117 Heidelberg, Germany
   \and
   Department of Physics and Electronics, Christ University, Hosur Road, Bangalore 560029, India
   \and
    Inter-University Centre for Astronomy and Astrophysics, Post Bag 4, Pune 411007, India
    \and
    Wits Centre for Astrophysics, School of Physics, University of the Witwatersrand, 1 Jan Smuts Avenue, Johannesburg, 2000, South Africa
    \and
    Department of Astronomy, University of Cape Town, Private Bag X3, Rondebosch 7701, South Africa
    \\
    \email{gourab.giri@up.ac.za}
             }

   \authorrunning{G. Giri et al.}
 
  \abstract
   {The persistence of radiative signatures in giant radio galaxies (GRGs $\gtrsim 700$ kpc) remains a frontier topic of research, with contemporary telescopes revealing intricate features that require investigation.}
   {This study aims to examine the emission characteristics of simulated GRGs, and correlate them with their underlying three-dimensional dynamical properties.}
   {Sky-projected continuum and polarization maps at 1 GHz were computed from five 3D relativistic magnetohydrodynamical (RMHD) simulations by integrating the synthesized emissivity data along the line of sight, with the integration path chosen to reflect the GRG evolution in the sky plane. The emissivities were derived from these RMHD simulations, featuring FR-I and FR-II jets injected from different locations of the large-scale environment, with propagation along varying jet frustration paths.}
   {Morphologies, such as widened lobes from low-power jets and collimated flows from high-power jets, are strongly shaped by the triaxiality of the environment, resulting in features like wings and asymmetric cocoons, thereby making morphology a crucial indicator of GRG formation mechanisms. The decollimation of the bulk flow in GRG jets gives rise to intricate cocoon features, most notably filamentary structures—magnetically dominated threads with lifespans of a few Myr. High-jet-power cases frequently display enhanced emission zones at mid-cocoon distances (alongside warmspots around the jet-head), contradicting the interpretations of the GRG as a restarting source. In such cases, examining the lateral intensity variation of the cocoon may reveal the source's state, with a gradual decrease in emission suggesting a low-active stage. This study highlights that applying a simple radio power–jet power relation to a statistical GRG sample is unfeasible, as it depends on growth conditions of individual GRGs. Effects such as inverse-Compton CMB cooling and matter entrainment significantly impact the long-term emission persistence of GRGs. The diminishing fractional polarization with GRG evolution reflects increasing turbulence, underscoring the importance of modeling this characteristic further, particularly for even larger-scaled sources.}
   {}

   \keywords{Galaxies: active -- Galaxies: jets -- galaxies: groups: general -- Magnetohydrodynamics (MHD) -- Methods: numerical }

   \maketitle
%

\section{Introduction}
Extragalactic jets emerging from actively accreting supermassive black holes (SMBHs) located in galactic nuclei (active galactic nuclei: AGN), have been traced through their radio synchrotron emissions since the 1950s \citep{Jennison1953,Hazard1963}. Currently, such AGN outflows are being widely investigated through multi-wavelength and multi-messenger approaches, with the origins of such jets linked to the interplay between magnetic fields and rotation—either of the SMBH or of the accretion disk through which matter is funneled into the SMBH \citep[cf.][]{Blandford2019}. The multi-wavelength properties of jets are now generally understood to be primarily governed by non-thermal processes like synchrotron and inverse-Compton mechanisms \citep[cf.][]{Hardcastle2020}. However, these observed emission processes are greatly influenced by a complex set of intrinsic microphysical processes (e.g., particle re-acceleration, matter entrainment, turbulence), many of which are still highly debated in astrophysical flows \citep[cf.][]{Peer2014}.

With recent advancements in radio observations using high-resolution and sensitive telescopes, there is now a promising expansion in exploring the lower-energy hump of the jets' emission spectra dominated by synchrotron mechanism \citep[cf.][]{Saikia2022,Mahatma2023}. The observations are not only detecting new features associated with jet evolutions \citep{Ramatsoku2020,Brienza2021,Bruggen2021,McKinley:2021,Knowles2022,Ubertosi:2025}, but also revealing detailed insights into the micro-scale processes governing such emission characteristics \citep{Turner2018,Mahatma2020,Condon2021}. In parallel with these investigations, ongoing advancements in numerical techniques (e.g., the Lagrangian-Eulerian particle-fluid technique) have paved the way for testing these intricate processes in high-resolution simulation grids, thereby enhancing our understanding of the impact of modeling such phenomena \citep{Vaidya2018,Winner2019,Ogrodnik2021,Dubey2023,Dubey2024}.

With these progresses, longstanding inquiries regarding radio galaxies have been revitalized, particularly the question of how far AGN jets can travel after their generation at the launching sites. Recent discoveries of distinct extragalactic jets extending nearly 5 and 7 Mpc in linear sizes necessitate a closer examination of their evolution mechanisms \citep{Machalski2008,Oei2022_5Mpc,Oei2024_7mpc}, especially when compared to the existence of AGN jets that appear to saturate in their evolution on galactic scales \citep{Odea2021,Baldi2023}. The question is particularly relevant given the energy-dissipating effects of processes such as radiative and adiabatic cooling, along with the entrainment of ambient matter, on the radiating non-thermal particles of these sources across different redshifts \citep{Odea1997,Schoenmakers2000,Rossi2008}. To further complicate the context, the detection of persistent emission in the radio lobes of distinct `giant' radio galaxies, which have ceased their active phase by halting jet ejection \citep{Subrahmanyan2008,2014ApJ...788..174B,Cotton2020,Bagchi2024}, calls for studies that explore the correlation between the dynamical evolution of these sources and their conversion into radiative fingerprints.

Giant Radio Galaxies (GRGs) are a minor subclass of radio galaxy population characterized by their extensive linear sizes, reaching 700 kpc or more with the current Hubble constant value, $H_0 = 70\, {\rm km\, s^{-1}\, Mpc^{-1}}$ \citep[some studies suggest a revised threshold of 1 Mpc, cf.,][]{Dabhade2023}. Alongside efforts to link morphological traits with hypothesized formation mechanisms \citep{Subrahmanyan1996,Malarecki2015,Cotton2020}, current research on GRGs aims to reconstruct their intrinsic three-dimensional dynamical properties from their projected emission maps and assess the feasibility of the observed states within the framework of established jet evolution theories \citep{Oei2023_length_dist}. For instance, properties like the source's spectral behavior and the magnetic-field estimates provide insights into the evolutionary timescales of jet-cocoon structures \citep{Machalski2009,Machalski2011}. Meanwhile, the fractional polarization values and the magnetic-field distribution reveal zones of internal and external stress layers, highlighting regions of shear and compression within and in the vicinity of the cocoon \citep{Laing1981, Mack1997, Cotton2020}. Studies on Faraday Rotation Measure emphasize the value of using GRGs as probes of estimating ambient-particle density and the correlation length of small-scale environmental turbulence \citep{Cantwell2020,Stuardi2020}. A key finding, as illustrated in \citet{Stuardi2020}, suggests that larger GRGs are more likely to be detected in polarized emission, indicating that their extensive lobes extend beyond the dense central environment and evolve within a rarified medium in the peripheral regions. This observation is also consistent with several continuum studies of GRGs \citep{Subrahmanyan2008,Oei2024_filament}, highlighting their prevalence within the relatively low-density warm-hot intergalactic medium \citep[WHIM;][]{Dave2001}, typically situated beyond the typical boundaries of galaxy groups or clusters.

Despite extensive numerical efforts to simulate AGN jets including the effects of microphysical processes like particle re-acceleration and cooling \citep{Borse2021,Mukherjee2021,Kundu2022,Dubey2023}, no such simulations to date have specifically addressed giant radio galaxies. This gap is largely due to the substantial computational demands required to resolve small-scale physics ($\sim 100$s of pc) while allowing jets to evolve over extended scales of a few megaparsec. In fact, numerical modeling to test the genesis hypotheses of GRGs (i.e. recognizing the dominant factors contributing to their growth: (a) longer temporal evolution, (b) efficient, collimated jets, (c) low-density environment to jet propagation, or (d) rapidly propagating restarting jets) has been notably scarce until recently. A 2D axisymmetric model has been introduced recently to explore the evolution of symmetric jets within a dense cluster environment, offering a preliminary framework for studying a subset of GRGs \citep{Duan2024}. However, it is now well established that the imposed symmetry and reduced dimensionality in these models often lead to discrepancies in results when compared to more accurate 3D simulations \citep{Mignone2010,Bodo2016,Massaglia2016,Kundu2022}.

Recognizing the necessity of investigating the influence of factors in producing extended GRGs ($\gtrsim 1$ Mpc) within a full 3D framework, only recently has a set of five key models been developed \citep{Giri2025}, spanning a diverse range of jet-environment configurations in relativistic magnetohydrodynamic (RMHD) settings. These models employ symmetry-breaking mechanisms intrinsic to three dimensional configurations, designed to more closely mimic realistic jet-medium interactions. The resulting dynamical features and their evolutionary implications are documented in \citet{Giri2025} (hereafter referred to as \citetalias{Giri2025}). While this study marks the first to develop 3D RMHD models for GRGs, the resolution achieved was necessarily limited. The entire jet injection region is defined over a cylindrical volume of 10 cells, with the jet diameter corresponding to 2 compute cells—a compromise to balance physical accuracy and computational efficiency at this stage. Nonetheless, efforts were made to minimize the impact of this constraint, particularly by leveraging the natural expansion of the jet following its injection. A detailed discussion of these mitigation strategies and the effects of resolution enhancement can be found in \citetalias{Giri2025}. 

Given the significance of transforming 3D dynamical data into insightful 2D projected emission maps, as discussed earlier, we intend to spearhead the radiative modeling of simulated GRGs. This study aims to comprehensively examine the correlation between the intrinsic dynamical behavior of (giant) jetted sources and their continuum and polarization properties. Additionally, it explores the underlying physical mechanisms that contribute to the suppression of the radiative signatures from such sources.

This paper is structured as follows: In Section~\ref{Sec: (Extended) insights on Paper I}, we provide a concise overview of the relevant details and conclusions from \citetalias{Giri2025} that form the foundation of this study. Section~\ref{Sec:Formulating radiative maps for the present investigation} presents the transformation of dynamic maps from \citetalias{Giri2025} into their corresponding radiative signatures at 1 GHz, which are examined in detail in this paper. This is followed by a discussion in Section~\ref{Sec:Exploring velocity patterns in GRGs} on the matter-transport processes responsible for generating the GRG cocoon. In Section~\ref{Sec:Radio continuum maps at 1 GHz}, we explore the emission characteristics of the simulated GRGs, providing several quantitative insights into their properties. Section~\ref{Sec:Polarization measures at 1 GHz} examines the polarization characteristics of the simulated structures, linking them to their underlying dynamical properties. We present a summary of our key findings in Section~\ref{Sec:Summary}.

\section{Insights from previous work} \label{Sec: (Extended) insights on Paper I}
This section provides a concise overview of the simulation framework and results from \citetalias{Giri2025}, revisiting its numerical formulation designed to test the GRG hypothesis and summarizing key findings that underpin the present analysis.

\subsection{Revisiting simulation setup of Paper I}\label{Sec:Revisiting Paper I}
The five 3D simulations presented in \citetalias{Giri2025} feature an ambient medium characterized by a King's $\beta$-profile, 
with the selected parameter values ensuring a stable galaxy group medium, thus maintaining a balance between gravity and pressure forces.
In Cartesian coordinates, the density follows a profile
\begin{align}\label{Eq:Ambient_medium}
    \rho (x',\, y',\, z) &= \rho_{0} \left(1+\frac{(x'-x_0)^{2}}{a^{2}}+\frac{(y'-y_0)^{2}}{b^{2}}+\frac{(z-z_0)^{2}}{c^{2}}\right)^{-3\beta/2}\\
    x' &\equiv x\,{\rm{cos}}10^{\circ} - y\,{\rm{sin}}10^{\circ}\\ 
    y' &\equiv x\,{\rm{sin}}10^{\circ} + y\,{\rm{cos}}10^{\circ}
\end{align}
where, $\rho_0$ represents the core density (0.001 amu/cc, i.e., in atomic mass / cm$^3$), 
while $x'$ and $y'$ denote axes rotated by $10^\circ$ in order to break the symmetry of the jet flow within the ambient medium
(insights on other parameters reveal as we discuss further).
In all simulations, a cylindrical under-dense jet with a density of \(\rho_j = 10^{-5} \times \rho_0\), a radius of 1 kpc (corresponding to the resolution of the computational domain), and an initial length of 3 kpc is injected into the medium. The setup considers one-sided jet injection and propagation along the negative \(x\)-direction within a simulation domain extending $\sim 700$ kpc in the jet-flow direction. 

The choice of a 10 degree offset (in Eq.~\ref{Eq:Ambient_medium}) is introduced to mimic jet's propagation in a more realistic environmental configuration rather than a perfect symmetry. Immediately after jet injection (after 3 kpc; length of the injection cylinder), the jet encounters a medium whose major/minor axis is misaligned with the jet-flow axis by approximately one cell at 5 kpc—consistent with the resolution of the
computational grid. This asymmetry becomes increasingly significant with distance, reaching offsets of $\sim 9$ cells at $50$ kpc and $\sim 44$ cells at $250$ kpc.

The only varying conditions across the five runs are (a) the configuration of the ambient medium, which is tri-axial, with jets propagating either from its center or its edges, and (b) the classification of the jets into two types based on power, \citet{Fanaroff1974} I and II types, injected into the computational domain.
This formulation has been designed to test two hypothesized models for GRG formation \citep{Dabhade2023}:
one suggesting that the dominant factors contributing to their growth is the more rarefied ambient medium, and the other proposing
that the high jet-power characteristics may have enabled the jet to overcome the frustration imposed 
by the surrounding environment.

\begin{figure*}
\centering
\includegraphics[width=2\columnwidth]{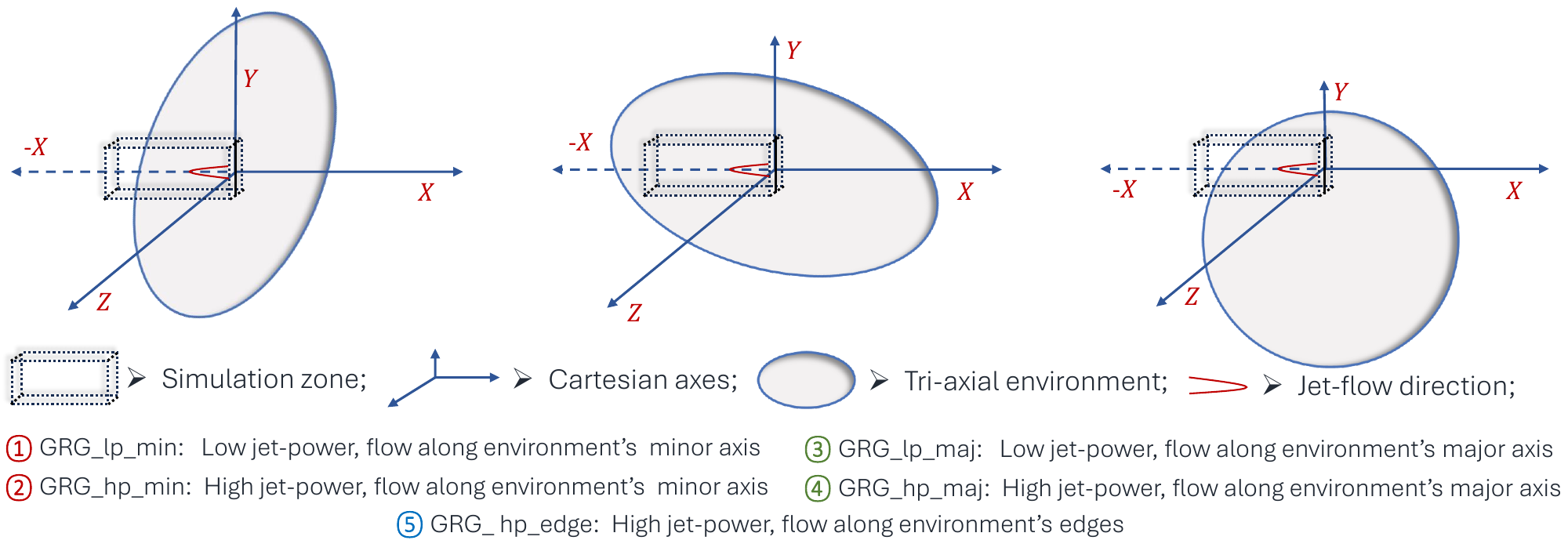}
\caption{Schematic representation of the jet-environment interaction framework developed to investigate the influence of factors on the development of giant radio galaxies. The setup simulates a one-sided relativistic jet propagating up to $\sim 700$ kpc through a stratified medium representing a galaxy-group environment, capturing the varying influence of the environment on the jet's motion. We also used two different jet powers to mimic Fanaroff-Riley type I and II jets, allowing us to examine the role of the jet's primary thrust in the development of such radio galaxies. This parameter space yields a total of five simulation scenarios, discussed in detail in \citetalias{Giri2025} and also concisely in Section~\ref{Sec:Revisiting Paper I}.}
\label{Fig:Setup} 
\end{figure*}

A schematic diagram illustrating the five models is presented in Fig.~\ref{Fig:Setup}. 
The simulations are labeled such that their nomenclature indicates the direction of the jet flow within the ambient environment
and the associated jet power.\footnote{\label{fn:Sim_Cases} Five simulations correspond to:
\begin{enumerate}[left=0pt]
\renewcommand{\labelenumi}{\Roman{enumi}.}
\item "GRG\_lp\_min": FR-I jet injected from the center, propagating along minor axis of a tri-axial medium, 
\item "GRG\_hp\_min": FR-II jet injected from the center, propagating along minor axis of a tri-axial medium,
\item "GRG\_lp\_maj": FR-I jet injected from the center, propagating along major axis of a tri-axial medium,
\item "GRG\_hp\_maj": FR-II jet injected from the center, propagating along major axis of a tri-axial medium,
\item "GRG\_hp\_edge": FR-II jet injected from the edges of the tri-axial medium and propagating along its boundary.
\end{enumerate}
}
We inject jets with low power, $L_j = 2.3 \times 10^{44}$ erg/s, denoted by `lp' (Lorentz factor 3),
and jets with high power, $L_j = 7.2 \times 10^{44}$ erg/s, denoted by `hp' (Lorentz factor 5), 
corresponding to powerful FR-I jets and typical FR-II jets, respectively \citep{Schoenmakers2000,Machalski2008}.
The acronyms `min', `maj', and `edge' refer to the jet's injection position and flow direction within the tri-axial environment. Specifically, `min' indicates a jet originating at the center and propagating along the minor axis (encountering less obstacles in its path), `maj' represents a jet flowing towards the major axis (facing significant obstacles in its path), and `edge' denotes a jet injected from and propagating along the edges ($x_0 = 0$, $z_0 = 0$, $y_0 = -6$; referring jet's injection at 600 kpc from the center) of the tri-axial medium (experiencing minimal obstacles in its path). The tri-axiality and axes configuration of the ambient medium are characterized by the parameters $a$, $b$, and $c$ (the effective core radii), with values varying between 33 kpc and 66 kpc (`min': $a,\, c \equiv 33$ kpc and $b = 66$ kpc, `maj': $b,\, c \equiv 33$ kpc and $a = 66$ kpc, `edge': $a,\, b,\, c \equiv 66$ kpc; refer to \citetalias{Giri2025}).

Fig.~\ref{Fig:Setup} summarizes our simulation runs by a schematic representation, 
highlighting the environmental structure, the direction of the jet flow, 
the simulation domain, the configuration of the coordinate axes, 
all for the five simulation setups considered. 

The simulations were performed in dimensionless units.
A conversion to physical units can be conducted based on three fundamental scales:
the length, $L_0 = 100$ kpc, the velocity, $V_0 = c$, i.e. the speed of light, 
and the density, $\rho_0 = 0.001$ amu/cc.

Other units can be derived using these fundamental scales, including 
the time, $T_0 = L_0/V_0 = 0.326$ Myr, 
the gas pressure, $P_0 = \rho_0 V_0^2 = 1.5 \times 10^{-6}$ dyn/cm$^2$, 
and the magnetic field strength, $B_0 = \sqrt{4\pi\rho_0V_0^2} = 4.3 \times 10^{-3}$ Gauss.

On all sides of the simulation domain an outflow boundary condition is applied, 
except for the boundary layer hosting the jet injection nozzle (right boundary, along $x-$axis; Fig.~\ref{Fig:Setup}),
where we assign a reflective boundary condition. 
This setup is meant to mimic one half of a bipolar jet simulation, 
in particular for cases for which a substantial amount of backflow material 
reaches the jet injection region. The outflow boundaries apply a zero-gradient extrapolation which allows matter to freely
leave the domain but does not artificially suppress inflow velocities. At earlier times, when the jet is still propagating
well within the simulation volume, the surrounding environment near the boundaries of the
simulation box remains effectively static, as governed by the combined balance of pressure
and gravity set in the initial conditions for the ambient medium.

It is important to note that in all cases the jet is injected (along negative $x-$axis) 
with a magnetic field in a purely toroidal configuration, 
which is defined as located in the plane perpendicular to the jet flow 
direction (i.e., in the $y-z$ plane),
\begin{equation}
    B_{y} = B_{j}\, r\, {\rm sin(\theta)},\ \ 
    B_{z} = -B_{j}\, r\, {\rm cos(\theta)}
\end{equation}
where, $(r,\, \theta)$ is the polar coordinate in the $(y,\, z)$ plane. 

The strength of the injected magnetic field ($B_j$) is calibrated by assuming that the 
magnetic energy constitutes $\sim 1$\% of the enthalpy of the jet beam while injection \citep{Mukherjee2020,Rossi2017}. 
This ratio is referred to as $\sigma$, usually denoted as magnetization,
and is defined as,
\begin{equation}
    \sigma = \frac{B_j^2}{\gamma^2 \rho_j h_j};\,\,\,\, \rho_j h_j = \frac{5}{2}P_j + \sqrt{\frac{9}{4}P_j^2 +(\rho_j)^2}
\end{equation}
where 
$\rho_j h_j$ is the matter-energy flux (enthalpy), 
$P_j$ is the pressure,
and $\gamma$ the bulk Lorentz factor of the jet (subscript $j$ for jet). 
Notably, the formulation of $\sigma$ is based on the Taub-Matthews equation of state, 
commonly used to describe a relativistic gas \citep{Taub1948,Mignone2005}.
We choose, $\sigma = 0.01$, implying $\langle B_j \rangle \simeq 9\,\mu$G 
at the injection distance from the galactic center of 3 kpc.

We apply two tracer variables in order to be able to disentangle between material of 
the ambient medium ($Q_1$) and the jet ($Q_2$).
Tracers are passive scalar variables that evolve according to the advection equation
\citep{Matthews2019},
\begin{equation}
    \frac{\partial(\rho Q_i)}{\partial t} + \nabla \cdot (\rho Q_i v) = 0
\end{equation}
When tied to a fluid as an initial condition, their subsequent evolution is governed by 
the density and velocity, $\rho,\, v$, of the associated fluid. 
This approach allows us not only to track the individual evolution of each fluid in a 
multi-fluid system, but also to determine the relative fraction of each fluid within a 
computational cell (value varying between 0 to 1). 
We emphasize, that this capability is essential for analyzing the matter entrainment 
processes that crucially influence the radiative appearance of GRGs.

\subsection{Revisiting relevant highlights of Paper I}
For the reader's convenience we repeat the following key conclusions from \citetalias{Giri2025}, as they are exceedingly relevant in the context of our 
current study.
\begin{enumerate}
    \item We observed the emergence of distinct cocoon morphologies under different scenarios, evolving over varying dynamical timescales with jets of different power. This led to the conclusion that a combined information of morphology, jet power, and structural age of GRGs holds significant potential to unravel the underlying formation mechanisms.  
    \item The simulated lobes of giant radio galaxies are over-pressured across all models, indicating that lobe expansion remains active as long as the jet activity persists. The lateral (to jet-flow) variation in pressure showed a sharp jump in values at the cocoon-ambient medium interface irrespective of the models, following the overpressure nature of the lobes.
    \item The axial ratio, defined as the ratio of cocoon length to width, reveals that powerful jets exhibit rapid, elongated expansion with arrowhead-like structures. In contrast, low-powered jets or jets encountering stronger environmental interaction produce wider lobes.
\end{enumerate}
We will keep these key insights in mind while further investigating the emission behavior of such radio galaxies and will return to these points when searching for correlations between the 3D dynamical evolution and the projected 2D emission maps. 

\section{Formulating radiative maps for the present investigation} \label{Sec:Formulating radiative maps for the present investigation}
Here, we outline the techniques adopted to convert the dynamical state of the simulated GRGs, presented in \citetalias{Giri2025}, into emission and polarization maps for further examination.

\subsection{Post-processing I: Generating intensity maps} \label{Sec:Post-processing I: Generating intensity maps}
We have converted the three-dimensional data of the simulated structures corresponding to the five simulation cases, as obtained at the evolved epochs detailed in \citetalias{Giri2025}, (see also Fig.~\ref{Fig:Velocity_dynamics}, for a reference), into informative projected emission maps.

To evaluate the synchrotron emissivity of the simulated GRGs, we build upon the framework outlined in \citet{Vaidya2018}, while adopting a more simplified dynamic-to-emission conversion methodology as described in \citet{Meenakshi2023}. The foundational approach employed here excludes the effects of non-thermal particle re-acceleration and radiative cooling but accounts for naturally evolving dynamical processes like matter entrainment and adiabatic cooling. The choice of this methodology, i.e. adopting a simplified form of Bessel function in emissivity evaluation, stems from the resolution requirements necessary to accurately incorporate all micro-scale processes \citep{Dubey2023}, which are nearly unfeasible with most contemporary computational resources, particularly for giant radio galaxy simulations (as also discussed in detail in Appendix~A.1. of \citetalias{Giri2025}). This limitation has led to the recommendation of utilizing 2D models \citep{Kundu2022}. However, as noted earlier, reduced dimensionality and imposed symmetry often result in deviations from the outcomes of fully 3D models \citep[e.g.,][]{Mignone2010}, which justifies the adoption of this approach for the present study.

The emissivity equation for an ensemble of electrons (mass: $m_e$, charge: $e$) 
distributed in a power-law fashion over an energy spectrum ($E_i$ to $E_f$), defined through Lorentz factors of $\gamma_i\, (10^2)$ to $\gamma_f\, (10^6)$ with a fixed power-law index $p$ \citep[$= 2.2$;][]{Machalski2011}, can be expressed in co-moving frame as:
\begin{equation}
    \mathcal{J'}_{\rm syn} (\nu', \vec{\hat{n'}}, \vec{B'}) = \frac{\sqrt{3}e^3}{4\pi m_ec^2} \left| \vec{B'} \times \vec{\hat{n'}} \right| \int_{E_i}^{E_f} \mathcal{N'}(E') F(x) \, dE'
\end{equation}
where, $\nu'$ is the emitting frequency, and vectors $\vec{B'}$ and $\vec{\hat{n'}}$ are the system's magnetic field and the line of sight direction, respectively. Assuming isotropic conditions for the particle distribution $\mathcal{N'}(E')$, and incorporating the Bessel integrals for the Bessel function, $F(x)$ with
$x \equiv \nu'/\nu'_{cr}$, the ratio of the emitted to the critical synchrotron frequencies, 
the expression for the emissivity simplifies to
\begin{equation}
\begin{split}
    \mathcal{J'}_{\rm syn} = N_0 \frac{3^{p/2}e^2\nu'^{-(p -1)/2}\left| \vec{B'} \times \vec{\hat{n'}} \right|^{(p+1)/2}}{2c(p + 1)} \\
    \Gamma \left( \frac{p}{4} + \frac{19}{12} \right) \Gamma \left( \frac{p}{4} - \frac{1}{12} \right) \left( \frac{e}{2\pi m_e c}\right)^{(p+1)/2} 
\end{split}
\end{equation}
Here, $N_0$ (the constant in particle spectral distribution) can be derived by assuming that the energy stored in the non-thermal particles is equal to a fraction of energy carried by either magnetic field, $E_B$ \citep{Giri2022_S,Giri2022_X,Upreti2024}, or by the internal energy, $E_{\rm th}$ \citep{Mukherjee2021,Meenakshi2023}, of the simulated system. Adopting the later, $N_0$ can be expressed as follows,
\begin{equation}
    N_0 = \frac{\epsilon E_{\rm th}}{m_e c^2 \int_{\gamma_i}^{\gamma_f} \gamma^{(-p+1)}\, d\gamma}
\end{equation}
where, $\epsilon$ indicates the fraction of internal energy ($E_{\rm th} = {\rm gas \, pressure}/({\rm Adiabatic\, gas\, constant} - 1)$) carried by the non-thermal particles; adopted to be 0.1, following \citet{Meenakshi2023}.

We note that all primed variables are defined in the co-moving frame and require conversion to the observer's frame as follows,
\begin{align}
    \mathcal{J}_{\rm syn} (\nu, \vec{\hat{n}}, \vec{B}) &= \mathcal{D}^2 \mathcal{J'}_{\rm syn} (\nu', \vec{\hat{n'}}, \vec{B'})\\
    \nu' &= \frac{1}{\mathcal{D}} \nu \\
    \vec{\hat{n'}} &= \mathcal{D} \left[ \vec{\hat{n}} + \left( \frac{\gamma^2}{\gamma+1} \vec{\beta}\cdot \vec{\hat{n}} -\gamma \right) \vec{\beta} \right]\\
    \vec{B'} &= \frac{1}{\gamma} \left[ \vec{B} + \frac{\gamma^2}{\gamma+1} (\vec{\beta} \cdot \vec{B}) \vec{\beta} \right]
\end{align}
where $\mathcal{D}$ is the Doppler factor and $\vec{\beta}$ is the velocity vector in units of speed of light. 
We thereby generate intensity maps ($I_{\nu}$) at a frequency of 1 GHz, which is relevant for several contemporary radio telescopes, by integrating the emissivity-data for the line of sight along the $z$-axis, 
in order to capture the giant radio galaxy's behavior in the plane of the sky, as has been the case for a majority of GRGs \citep{Oei2023_length_dist}. 

\subsection{Post-processing II: Generating polarization maps}\label{Sec:Post-processing II: Generating polarization maps}
Building on the definitions defined in Section~\ref{Sec:Post-processing I: Generating intensity maps}, we utilize the expression for polarized emissivity as reported in \citet{Vaidya2018},
\begin{equation}
    \mathcal{J'}_{\rm pol} (\nu', \vec{\hat{n'}}, \vec{B'}) = \frac{\sqrt{3}e^3}{4\pi m_ec^2} \left| \vec{B'} \times \vec{\hat{n'}} \right| \int_{E_i}^{E_f} \mathcal{N'}(E') G(x) \, dE'
\end{equation}
where, the Bessel function $G(x) = xK_{2/3}$. Utilizing the simplified form of the Bessel integral \citep{Rybicki1979}, the polarized emissivity in the observer's frame simplifies to \citep{Meenakshi2023},
\begin{equation}
\begin{split}
    \mathcal{J}_{\rm pol} = \mathcal{D}^2 N_0 \frac{3^{p/2}e^2\nu'^{-(p -1)/2}\left| \vec{B'} \times \vec{\hat{n'}} \right|^{(p+1)/2}}{8c} \\
    \Gamma \left( \frac{p}{4} + \frac{7}{12} \right) \Gamma \left( \frac{p}{4} - \frac{1}{12} \right) \left( \frac{e}{2\pi m_e c}\right)^{(p+1)/2} 
\end{split}
\end{equation}
where, the conversion of $\nu'$, $\vec{B'}$, $\vec{\hat{n'}}$ follows the equations highlighted in Section~\ref{Sec:Post-processing I: Generating intensity maps}. The Stokes parameters for linear polarization ($U_{\nu}$ and $Q_{\nu}$) can subsequently be computed along the line of sight ($z-$axis) as,
\begin{equation}
\begin{split}
    U_{\nu} (\nu, X, Y) &= \int_{-\infty}^{\infty} \mathcal{J}_{\rm pol}  \, \, \text{sin} 2\mathcal{X}  \, \, dZ \\
    Q_{\nu} (\nu, X, Y) &= \int_{-\infty}^{\infty} \mathcal{J}_{\rm pol}  \, \, \text{cos} 2\mathcal{X}  \, \, dZ
\end{split}
\end{equation}
where, the local polarization angle $\mathcal{X}$ is defined as \citep[following,][]{DelZanna2006}, 
\begin{equation}
\begin{split}
    \text{sin} 2\mathcal{X} = -\frac{2 q_X^{} q_Y^{}}{q_X^2 + q_Y^2}; \quad \text{cos} 2\mathcal{X} = \frac{q_X^2 - q_Y^2}{q_X^2 + q_Y^2}
\end{split}
\end{equation}
and,
\begin{equation}
\begin{split}
    q_X^{} = (1 - \beta_Z)B_X - \beta_X B_Z; \quad q_Y^{} = (1 - \beta_Z) B_Y - \beta_Y B_Z
\end{split}
\end{equation}

\noindent The $X$ and $Y$ axes formulate the sky-plane in the observer's frame, where the projected maps are generated. The angle $\mathcal{X}$ is measured clockwise from the $Y$-axis, given the axis is aligned towards the north in the sky-plane. Magnetic field lines are then overlaid on the Stokes-I image (intensity), with the polarization angle $\mathcal{X}$ being rotated by $90^\circ$. The length of the field lines is made proportional to the fractional polarization values, denoted as ${\scriptstyle \prod}$, at each point. The fractional polarization, ${\scriptstyle \prod}$, at the locations of synchrotron emissions is defined in terms of the Stokes parameters as,
\begin{equation}
    {\scriptstyle \prod} = \small \frac{\sqrt{U_{\nu}^2 \, \, + \, \, Q_{\nu}^2}}{I_{\nu}}
\end{equation}

\section{Exploring velocity patterns in GRGs}\label{Sec:Exploring velocity patterns in GRGs}
In this section, we extend the analysis from \citetalias{Giri2025} by examining the velocity patterns of the generated giant radio galaxy structures to understand how different model assumptions result in different cocoon structures.

\begin{figure*}
\centering
\includegraphics[width=2\columnwidth]{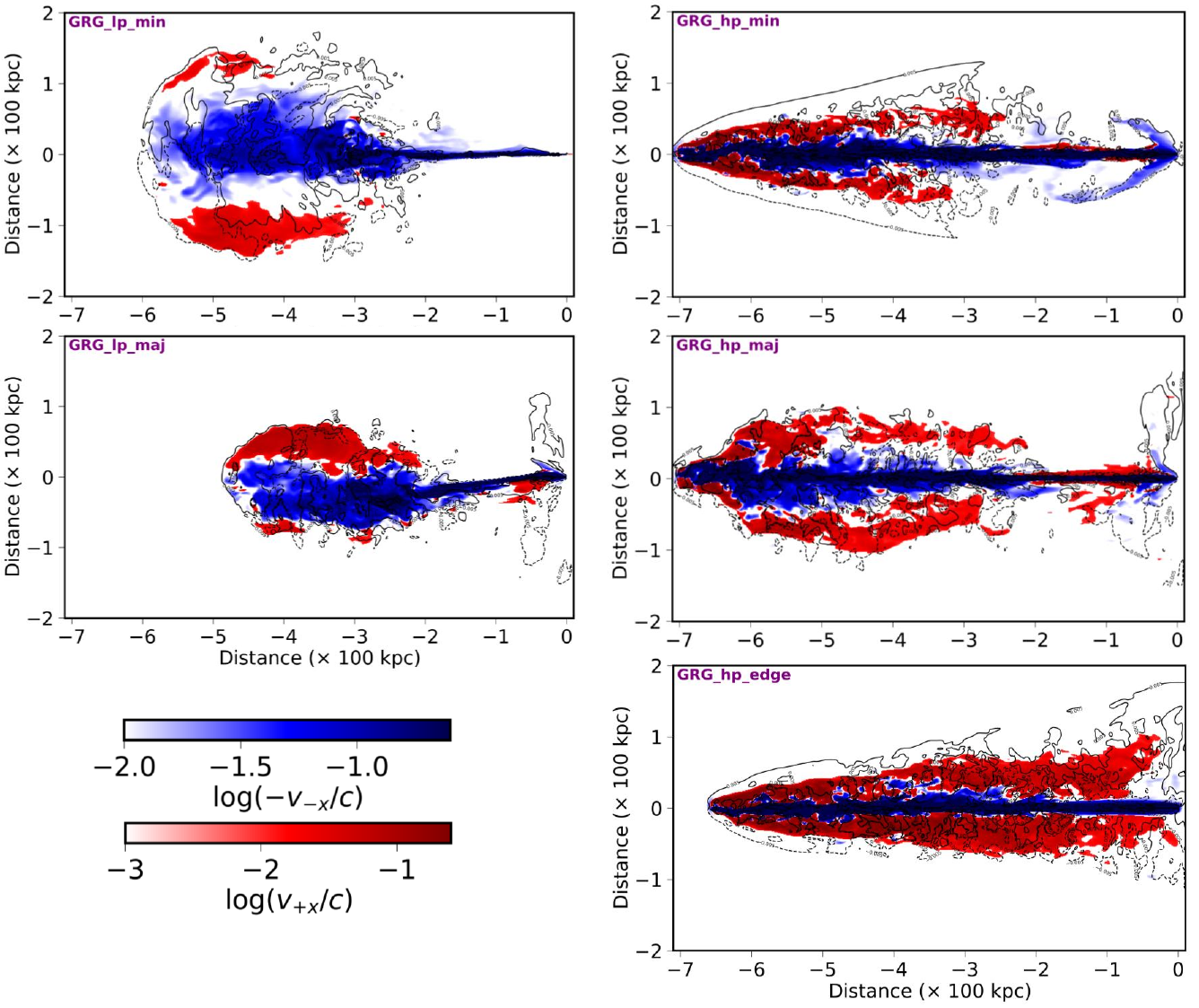}
\caption{Sliced velocity maps ($x-y$ plane, $z=0$) for the resultant giant radio galaxy morphologies from five simulation cases (labels indicated in each plot) are shown, illustrating the underlying matter-transport mechanisms responsible for these structures. These maps are displayed at dynamical ages of $\sim$ 167 Myr (GRG\_lp\_min), 196 Myr (GRG\_lp\_maj), 49 Myr (GRG\_hp\_edge), 137 Myr (GRG\_hp\_maj), and 69 Myr (GRG\_hp\_min), as also highlighted in \citetalias{Giri2025}. The colormap represents the $x-$component of velocity, capturing the bulk-flow (in blue) and back-flow (in red) dynamics, while the contours depict the $y-$component of velocity, with solid lines indicating $0.005c$ and dashed lines indicating $-0.005c$, to highlight the behavior of secondary structures forming within the domain. To analyze the impact of jet power on morphology and the influence of the tri-axial environment on the resulting structure, comparisons are made column-wise and row-wise, respectively.}
\label{Fig:Velocity_dynamics} 
\end{figure*}

The velocity plots shown in Fig.~\ref{Fig:Velocity_dynamics} are comprehensive yet straightforward to interpret. 
The $x-$component of velocity is represented in a colormap, where the $v_{-x}$ component, indicating the active jet's bulk 
flow, is shown in blue shades, while the backflowing matter is displayed in red shades via $v_{+x}$, 
both on a logarithmic scale for enhanced visualization. 

In addition, in order to capture the significance of matter flowing along the $y-$direction in certain cases—critical for revealing the full structural details—we have overlaid contours of the $y-$component of velocity at values of $0.005c$ ($v_{+y}$; solid lines) and $-0.005c$ ($v_{-y}$; dashed lines), where c being the light speed.

We may follow two approaches in order to comparing the distinct evolutionary
features of these runs, as we will discuss below.

\subsection{Impact of jet power}

\begin{figure}
\centering
\includegraphics[width=\columnwidth]{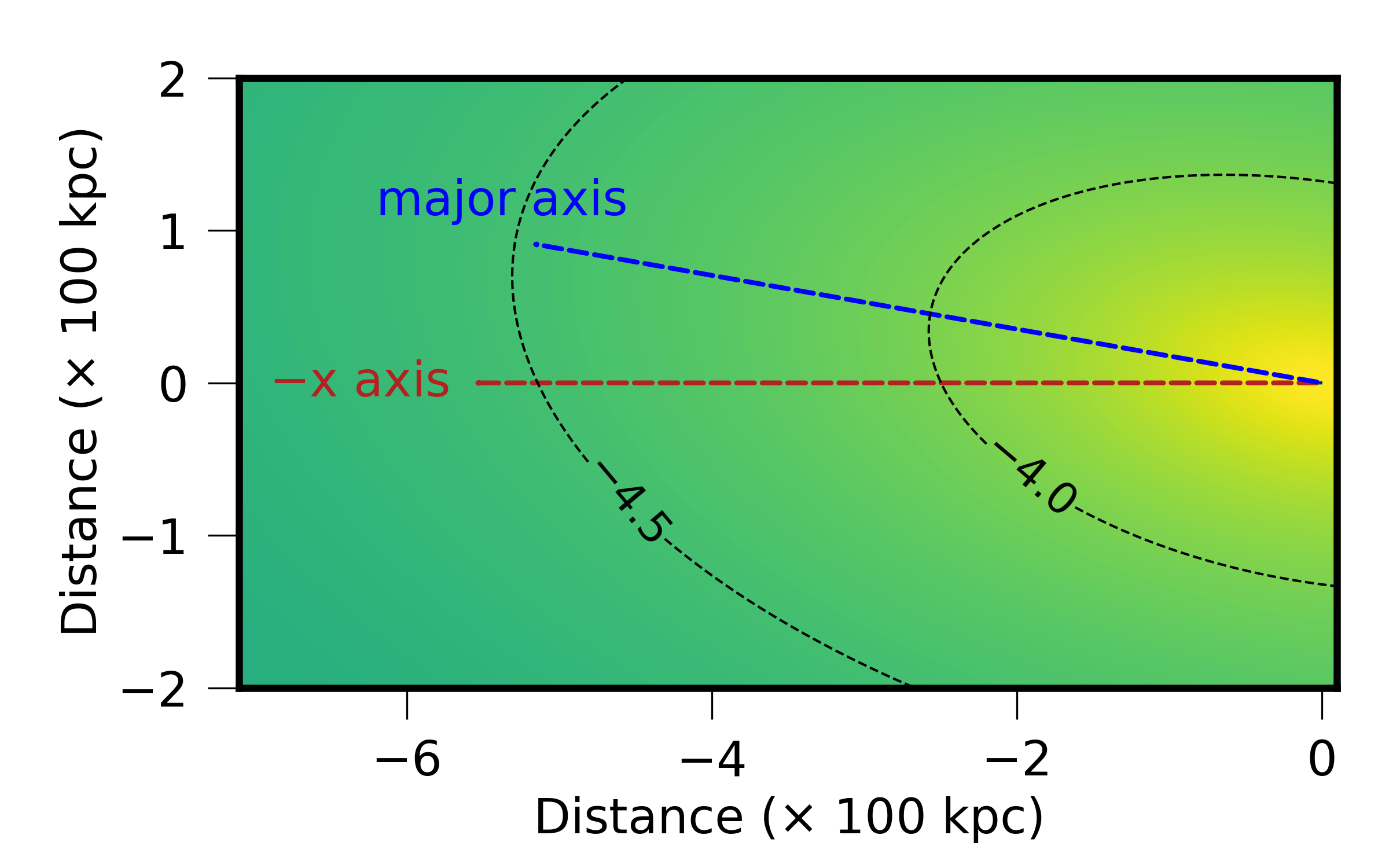}
\caption{Initial density configuration of the ambient environment, illustrating a $10^{\circ}$ rotation of the ambient medium's principal axis,
in order to introduce an asymmetry to the jet-flow, while the jet injection axis remains fixed along the negative x-direction. 
The density distribution is shown using colormap, with the corresponding $\log \rho [\rm amu/cc]$ values annotated along the contour lines. 
This represents one of the simulation scenarios employed, as shown in Fig.~\ref{Fig:Setup}.}
\label{Fig:Xaxis_Majoraxis} 
\end{figure}

A comparison between the low jet power models in the left column and the high jet power cases in the 
right column highlights how the differences in jet power may influence the general dynamical behavior. 
For example, we observe that lower-powered jets (FR-I like) tend to form wider lobe structures. 
This can be attributed to the jets' inability to maintain collimation over longer distances. 
This effect is further aggravated by slight bending (along $y-$axis) caused by the external ambient medium, as the medium is rotated by $10^{\circ}$ to break symmetry. An illustration of the initial configuration of the environment (for the `maj' cases) is presented in Fig.~\ref{Fig:Xaxis_Majoraxis}, highlighting the asymmetry introduced between the major axis and the jet propagation axis.

Together, these effects lead to a de-collimated, broad flow that distributes 
material over a wider opening angle \citep[similar to e.g.,][]{Monceau-Baroux2015,Nawaz2016}.
As a result of this divergence, the bulk flow speed of the fluid significantly decreases along the $-x$ axis.
This reduction in speed also limits the formation of backflow material, causing the material to propagate more slowly, which thereby struggles
to reach back the location where the jet originated. The combination of a wider bulk flow and a slowed backflow ultimately contributes to the formation of significantly wider lobes (refer to the $v_y$ contours). In this context, we note that the presence of mild back-flowing plasma in FR-I jets has been documented by \citet{Laing2012}, exhibiting characteristics consistent with our findings.

In contrast, the cases in the right column, characterized by higher flow speeds (FR-II like), 
consistently exhibit well-collimated jet beams up to the jet head. 
This indicates rapid propagation, as the jet thrust is concentrated over a smaller area \citep{Bromberg2011}.
Consequently, this results in a significant pressure imbalance at the jet head–ambient medium interface, 
generating a higher volume of backflowing material, even at greater speeds \citep{Cielo2017}. 

As a result, the backflowing material stretches out predominantly along the jet flow direction.
To note, the rapid expansion we observe for the high-powered jets is simply evident from the fact that they are able
to cover a linear extent of nearly 700 kpc within 50–140 Myr, 
whereas the low-powered jets reach a smaller linear extent of only up to 600 kpc, despite evolving over a 
longer duration of 170–200 Myr \citepalias{Giri2025}.

\subsection{Impact of the Environment}


We now investigate the influence of the environment. In Fig.~\ref{Fig:Velocity_dynamics}, this means comparing row-wise. We observe distinctions arising from the influence of the ambient medium's configuration. In the minor-axis cases shown in the top row, we observe that for a low-powered jet, decollimation leads to the lobe expanding laterally, aided by the pressure gradient force that is strongest along the minor axis \citep[similar to, e.g.,][]{Giri2023,Bruno2024}. As a result, the lobe's lateral size is largest in this case (GRG\_lp\_min) among all the models.

Conversely, for the high-powered case (GRG\_hp\_min), the jet remains collimated and maintains high thrust, enabling rapid propagation (again, aided by the pressure gradient). Consequently, the backflow material struggles to reach all the way back to the launching region, ultimately producing an arrowhead-shaped structure.

In the middle row (Fig.~\ref{Fig:Velocity_dynamics}), 
the jet propagates along the major axis of the environment. With greater obstruction to the jet flow due to the lower rate of change in pressure values (low pressure gradient) along this direction, the low jet power run exhibits lobes that are not as wide as those in the `GRG\_lp\_min' case. However, as the jet enters into the computing domain with the same power as `GRG\_lp\_min' and evolves for a longer period (196 Myr versus 167 Myr), subsequently failing to spread laterally, generates substantial backflow. 

This backflow returns to the zone of jet-injection and then follows the path of maximum pressure gradient, i.e., along the minor axis of the environment. This expansion of backflowing material in the shape of a wing is further captured in the $y-$component of velocity (Fig.~\ref{Fig:Velocity_dynamics}) and results in a structure known as X-shaped radio galaxies \citep[XRGs;][]{Giri2024_Review}. This mechanism of XRG formation is referred to as the classic backflow scenario \citep{Capetti2002,Rossi2017} and is believed to occur in many (giant) radio sources with X-shape \citep{Saripalli2009,Cotton2020,Bruni2021}.  

For high-powered jets flowing along the major axis, a similar X-shaped structure is formed. However, the prominence of the secondary lobes, or wings, is reduced since the jet propagates rapidly (less decollimation), leading to a decrease in the wing-to-lobe length ratio \citep[ratio < 0.8;][]{Cheung2007}. Notably, due to the jet's fast propagation, even though a significant amount of backflow is generated—as expected for higher-powered jets—the bulk fraction of matter struggles to return quickly to the jet injection zone. Consequently, a substantial amount of material accumulates within the active lobe cocoon (along $-x$ axis), creating a morphology known as a nose-cone shape \citep{STone2000}. 
The overlaid contours illustrate the $y-$velocity of gas-flow, showing explicitly the channeling of matter along the minor axis of the environment to form the wings. 

The asymmetry in the $v_y$ contours with respect to the jet-flow direction for `maj' cases can be attributed to the effect of tilted ambient medium, which is inclined by $10^{\circ}$ to break symmetry to the jet flow (Fig.~\ref{Fig:Xaxis_Majoraxis}). Furthermore, for the high-power cases, in general, the contour distribution reveals the formation of strong bow shocks in the environment due to the rapid expansion of the jet. In contrast, for the low-power cases, the bow shock is less prominent and remains almost attached to the entire frontal part of the lobe, reflecting slower progression of lobe expansion along the $x-$axis.

The bottom row highlights an intriguing case of GRG formation, where the jet flows along the edges of the environment with minimal obstacles, producing an arrow-head structure. The well-collimated jet propagates rapidly (covering $\sim$ 650 kpc in 49 Myr), generating substantial and fast backflows that return toward the jet origin. These backflows, influenced by buoyancy forces, create asymmetry in the cocoon's lateral extent as plasma follows the pressure gradient outward (from the centre of the ambient medium; Fig.~\ref{Fig:Setup}). This supplies a significant clue for guessing underlying GRG formation, as many are suspected to evolve outside galaxy groups or cluster media. Observationally, GRGs mapped near voids or along large-scale density gradients often show bending in their lobes \citep[e.g.,][]{Safouris2009,Malarecki2015}, consistent with this result.
\\

\noindent From these discussions, it is evident that the combined effects of the ambient medium and jet power drive the appearance of GRGs. Also noteworthy is the observation that in cases of high-powered jets, the colormap shows the presence of $v_{+x}$ components in the immediate vicinity of the jet beam, suggesting that material may have leaked from the jet-column due to instabilities, and subsequently flow alongside the jet. This is likely caused by the formation of a shearing layer between the sheath material and the jet-spine, which facilitates the leakage through Kelvin-Helmholtz (KH) instabilities \citep{Borse2021,Wang2023}. This phenomenon could be an intriguing factor for further investigation, as it may provide additional insights into the survival and stability of jet flows over even greater distances.

\section{Radio continuum maps at 1 GHz}\label{Sec:Radio continuum maps at 1 GHz}
This section provides a detailed analysis of the emission maps derived from the techniques outlined in Section~\ref{Sec:Formulating radiative maps for the present investigation}. It is divided into three subsections: the first (\S~\ref{Sec:Understanding GRG dynamics through emission maps}) discusses the morphological features observed in the maps for each GRG case; the second (\S~\ref{Sec: Quantitative exploration of the emission maps}) presents a quantitative analysis of these maps, leading to inferences relevant to observational studies; and the final subsection (\S~\ref{Sec:Correlating emission features with dynamical properties: Hotspot formation and filamentary extensions}) explores the correlation between the obtained radiative results and the underlying three-dimensional dynamical state.

\subsection{Understanding GRG dynamics through emission maps} \label{Sec:Understanding GRG dynamics through emission maps}
The generated emission maps effectively encapsulate the internal dynamical behavior of the GRG systems, presenting a more concise representation by reducing dimensionality. These maps integrate the combined influence of multiple three-dimensional properties into a single two-dimensional framework. Since the maps are integrated along the line-of-sight vector, any features distinctly observable in the emission maps indicate the prominence of those layers within the underlying 3D structure. Below, we discuss the projected sky maps classified according to their formation processes.

\subsubsection{Jet propagation through regions of moderate obstructions}
The 1 GHz emission maps shown in Fig.~\ref{Fig:Emission_minor}(A) reveal that extrinsic properties—such as the cocoon length, its geometric shape, and the overall projected axial ratio (length-to-width ratio)—exhibit consistent similarities with the dynamical maps \citepalias{Giri2025} for the chosen line-of-sight. Fig.~\ref{Fig:Emission_minor}(A) contains the evolved structures at ages of nearly 167 Myr and 69 Myr for a low-powered jet ("GRG\_lp\_min", top panel) and a  high-powered jet ("GRG\_hp\_min", bottom panel), respectively, propagating along the environment's minor axis direction.$^{\ref{fn:Sim_Cases}}$ 

\begin{figure}
\centering
\includegraphics[width=\columnwidth]{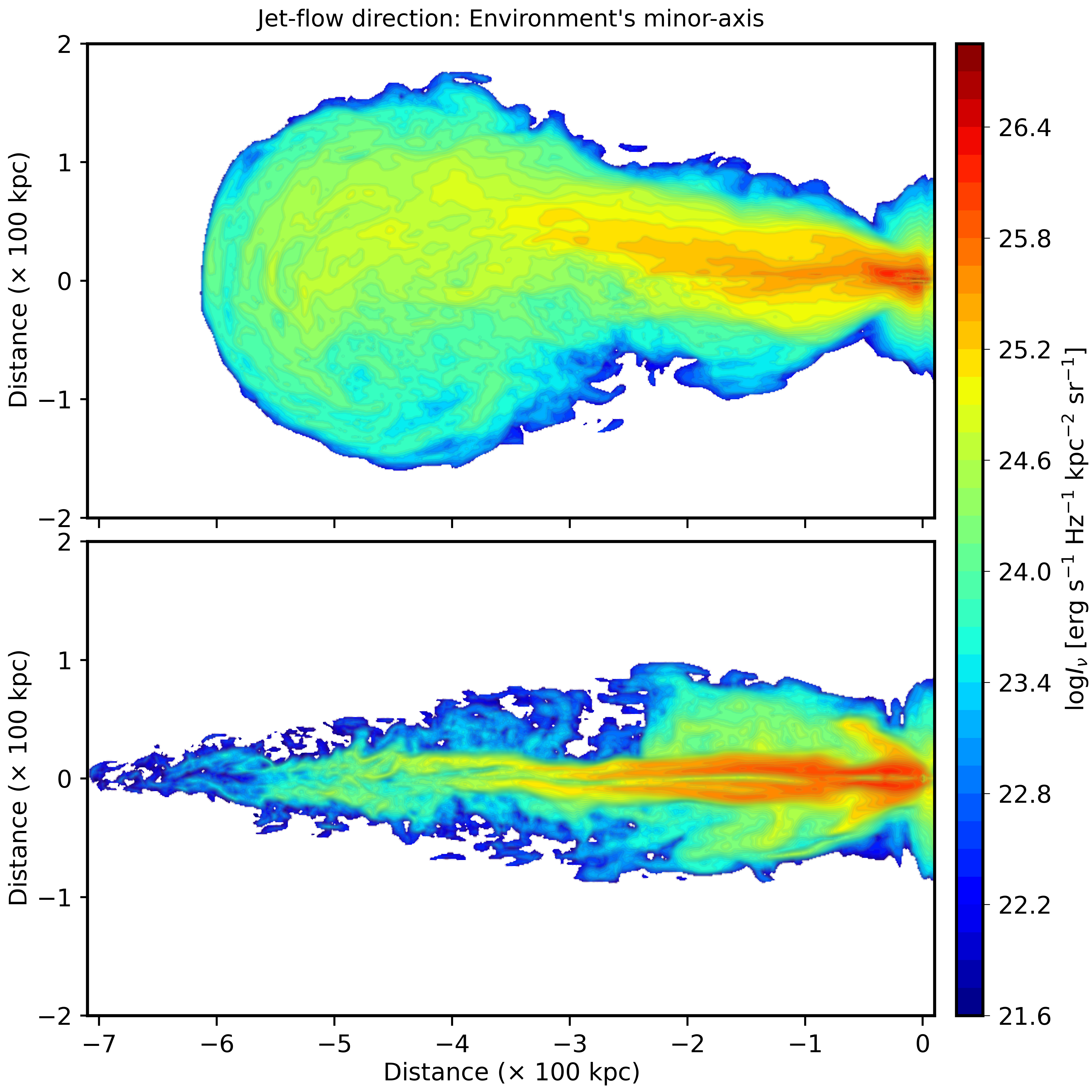}
\caption{A. Intensity maps (log $I_{\nu}$) at 1 GHz radio frequency are presented for two simulation cases: a low-power jet ("GRG\_lp\_min"; \textit{top}) and a high-power jet ("GRG\_hp\_min"; \textit{bottom}), both propagating along the minor axis of the ambient medium. While the overall structures resemble the dynamical maps from \citetalias{Giri2025}, the emission maps offer deeper insights, highlighting crucial aspects of the observability of these simulated giant radio structures at their evolved stages ($\sim$ 167 Myr: \textit{top}; $\sim$ 69 Myr: \textit{bottom}).}
\label{Fig:Emission_minor} 
\end{figure}

The decollimation of the jet flow, combined with the counterclockwise bending of the bulk flow, induces lobe formation in the "GRG\_lp\_min" case. This suggests that GRG lobes may not be passive regions of the jet-cocoon system but rather zones of intensified turbulence \citep[e.g.,][]{Kundu2022}, fostering the development of decollimated filamentary structures and cascading multi-scale vortices. The decollimated flow extends to the cocoon head along the linear growth direction, indicating ongoing lobe expansion.  This is corroborated by the presence of a mild bow shock adjoining the frontal lobe, as evidenced in the dynamical evolution outlined in \citetalias{Giri2025}.  The well-defined boundary of the lobe in this case is particularly notable, suggesting that, in addition to the structure being 1.9 times over-pressured relative to the surrounding medium, Kelvin-Helmholtz (KH) instabilities have a lesser impact on disrupting the cocoon in such low jet power scenarios \citep[low-magnetization cases;][]{Mukherjee2020}. 

The differences in cocoon disruption become particularly evident when comparing with the high-power jet case ("GRG\_hp\_min"). Despite having an over-pressure of four times that of the environment, the cocoon edges in this case are more susceptible to matter entrainment by KH instabilities. This phenomenon arises from the combination of high jet power and the substantial density contrast between the jet and the surrounding medium, which together ensure the survival of the jet spine but compromise the stability of the cocoon \citep{Rossi2008}. These findings underscore the critical role of instabilities and entrainment processes in shaping the jet-cocoon dynamics, particularly in the long-term evolution and survival of giant radio galaxies. 

In the "GRG\_hp\_min" case, the jet spine exhibits a bifurcated structure, attributed to the concentration of magnetic field lines along the jet beam's edges, followed by the formation of recollimation zones. This behavior aligns with high-resolution numerical studies \citep{Dubey2023} and is consistent with recent high-resolution, sensitive observations of extragalactic jetted radio sources \citep{Condon2021,Velovic2023,Wezgoweiec2024}. 


The "GRG\_hp\_min" case is particularly notable for two key reasons. (a) It exhibits the transition of the jet spine into a decollimated, turbulent state with tendril-like extensions, providing insight into the jet-lobe transition process in giant radio galaxies \citep[with relevance to smaller RGs, e.g.,][]{Condon2021,Velovic2023,Wezgoweiec2024}. (b) The decollimation of the jet spine at $\sim$ 450 kpc, despite the cocoon extending to 700 kpc, suggests that hotspots (or enhanced emission zones) located well within the cocoon of observed GRGs may not always indicate a restarting jet phase \citep[e.g.,][]{Sebastian2018,Andernach2021,Oei2022_5Mpc}. This observation may challenge the conventional interpretation of hotspot positions as definitive indicators of restarting activity in GRGs \citep[see the numerical models, e.g.:][]{Horton2023,Young2024}. Decollimation in this high-power jet case yields warmspot (wide emission zone) instead of a distinct hotspot, along with tendril-like filamentary structures, further analyzed in Section~\ref{Sec:Hotspots and jet dynamics} and \ref{Sec:Correlating emission features with dynamical properties: Hotspot formation and filamentary extensions}.

\subsubsection{Jet propagation through regions of maximal obstructions}\label{Sec:Jet propagation through regions of maximal obstructions}
Due to the jets propagating along a path with a lower pressure gradient (i.e., the major-axis direction) and encountering greater resistance to free flow in both the low jet power (top panel, Fig.~\ref{Fig:Emission_major}(B)) and high jet power cases (bottom panel, Fig.~\ref{Fig:Emission_major}(B)), the formation of lobe structures follows. This obstruction causes the jet to decelerate faster, resulting in earlier decollimation into tendril-like extensions compared to the jet flow along the minor axis situations, where collimation is preserved over a greater distance ($\sim$ 300 kpc vs. 400 kpc). Both scenarios exhibit noticeable wing emission, highlighting the influence of lateral deflection of backflow material from the active lobe. Additionally, the interaction of matter, likely originating from the opposite jet arm, contributes to this effect \citep[modeled using a "reflective" boundary condition near the jet injection boundary to replicate this natural phenomenon;][]{Capetti2002}. These effects cause an increase in internal energy within the wings highlighting the fact that wings are not passively evolving structures in XRGs. Instead, they function as regions of enhanced turbulence and random shock sites. This finding has important implications for understanding the persistent emission and anomalous spectral behavior observed in the wings of many XRGs, which is believed to be caused by particle re-energization, as indicated by both numerical studies and observations \citep{Giri2022_X,Gopal-Krishna2022,Patra2023}.

\begin{figure}
\ContinuedFloat
\centering
\includegraphics[width=\columnwidth]{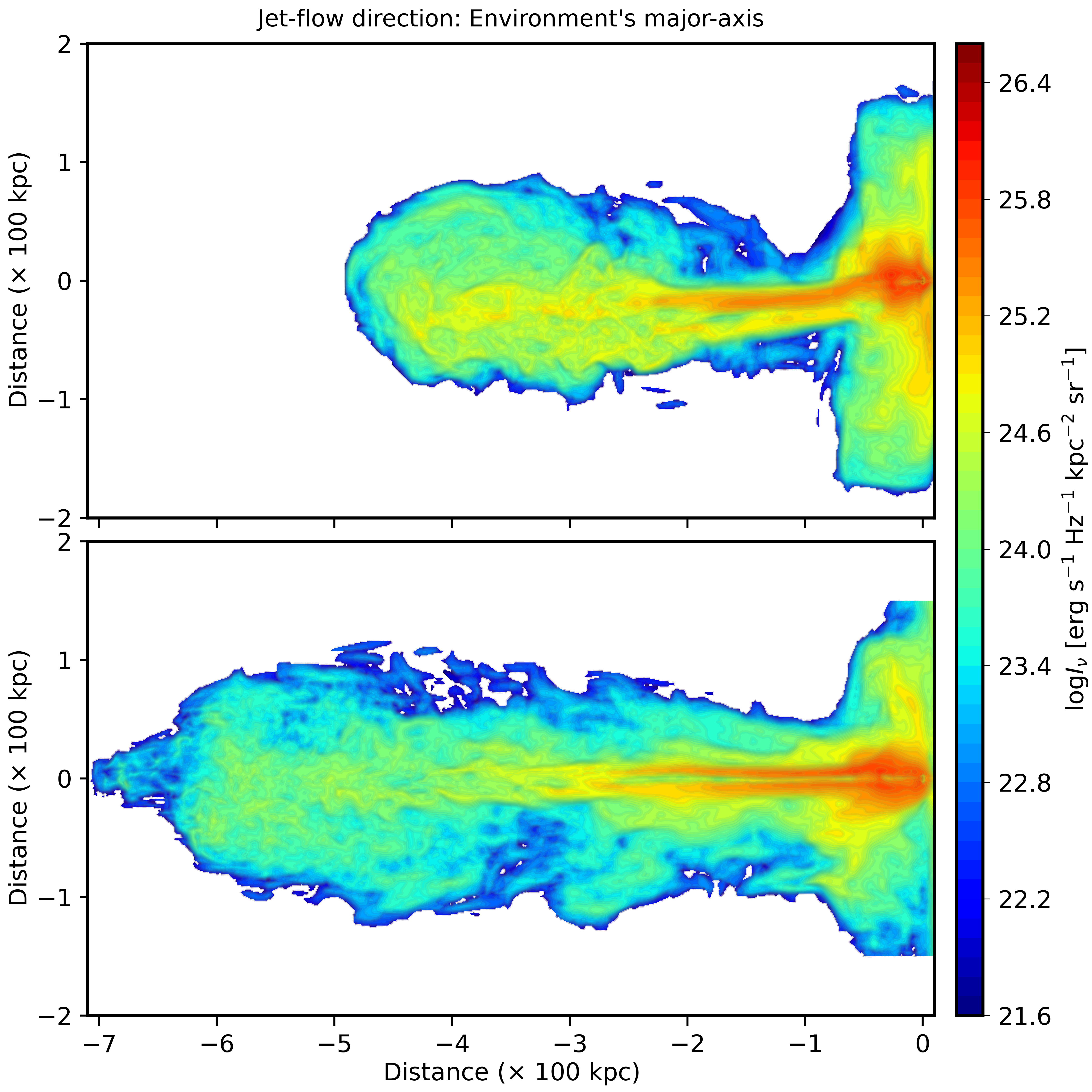}
\caption{B. \textit{Continued.} Intensity maps (log $I_{\nu}$) at 1 GHz radio frequency depicting the X-shaped radio galaxy in their giant stages, resulting from jet propagation along the direction of highest jet frustration (i.e., along the major axis direction). The low jet power case ("GRG\_lp\_maj"; \textit{top}) and the high jet power case ("GRG\_hp\_maj"; \textit{bottom}) exhibit distinct decollimation patterns, with both cases generating internal regions of emission valleys and enhanced emission ridges. The wings are luminous, indicating their active evolutionary stages ($\sim$ 196 Myr: \textit{top}; $\sim$ 137 Myr: \textit{bottom}). }
\label{Fig:Emission_major} 
\end{figure}

Focusing on the evolved structure (at $\sim$ 196 Myr) of the low jet power case ("GRG\_lp\_maj"), we observe notable behavioral features, such as the jet's bending in response to the ambient medium's $10^{\circ}$ rotational setup to break symmetry (Fig.~\ref{Fig:Xaxis_Majoraxis}). This interplay further induces the formation of a large vortex-like structure in the lobe, with jet material rotating in a clockwise direction (as also inferred from velocity streamlines; not shown here). Additionally, this motion leads to the creation of smaller-scale caldera-like structures, characterized by emission valleys surrounded by enhanced emission ridges. The compression of jet-matter (consequently the magnetic field; Section~\ref{Sec:Polarization measures at 1 GHz}) in the vicinity of lobe edges is particularly prominent \citep[e.g.,][]{Condon2021}. These dynamic processes ultimately contribute to the development of the lobe structure observed along the active jet axis.

In the high-powered jet case ("GRG\_hp\_maj"), the jet resists bending due to its higher thrust, and upon decollimation, the jet beam disperses the jet material over a wider angle forming spread-out emission ridges \citep[e.g.,][]{Cotton2020}. A slight asymmetry in the channeling of back-flow material has been observed in the lobe, with stronger channeling occurring towards the southern side with respect to the jet axis. Despite the jet's decollimation, the thrust appears to persist up to the frontal region, forming a top-hat-like structure (see, Section~\ref{Sec:Exploring velocity patterns in GRGs}). The sustained thrust of the jet results in an intensified interaction with the ambient medium in the top-hat region of the jet, which will be further discussed in Section~\ref{Sec:Hotspots and jet dynamics}. 

As the jet travels larger distance compared to "GRG\_lp\_maj", it loses internal energy and magnetic field strength more due to expansion, leading to a reduced contribution to the emission map from the frontal lobe part (compared to "GRG\_lp\_maj"; Fig.~\ref{Fig:Emission_major}(B)). This emission map  also reiterates that enhanced emission regions, deeply embedded within the structure, do not necessarily imply a jet restart \citep{Young2024,Fanaroff2021}. This underscores the need for additional diagnostics, such as pressure jumps at lobe boundaries (see Section~\ref{Sec:Insights from the lateral intensity variation}), to reliably differentiate between jet phases. The cocoon is susceptible to greater disruption in this case due to increased shear with the surrounding environment.

\subsubsection{Jet propagation through regions of minimal obstructions}
This scenario involves a high-powered jet propagating through the edges of the ambient medium ("GRG\_hp\_edge", Fig.~\ref{Fig:Emission_edge}(C)), where minimal flow resistance allows for rapid propagation, as evidenced by the jet covering $\sim 670$ kpc (one-sided) in $\sim$ $49$ Myr. At the same time, this also results in a higher shear between the cocoon and the surrounding environment, significantly affecting the visibility of the cocoon's extent. Due to the high thrust, the jet decollimates gradually (compared to "GRG\_hp\_maj"), generating finger-like filamentary extensions. Additionally, the increased jet speed leads to the leakage of matter from the jet spine, a feature observed as a result of enhanced shear between the jet sheath and spine, which fosters instability growth (see, Section~\ref{Sec:Exploring velocity patterns in GRGs}). The combination of these effects results in the mixing of matter from different regions, such as thermal and non-thermal particles, as well as freshly injected and aged particles \citep{Turner2018,Rossi2024}. Such mixed flow, along with backflow and buoyancy effects, creates greater turbulence within the cocoon, which explains the enhancement of emission closer to the jet-launching zone rather than near the jet head \citep[e.g., PKS 2356-61;][]{Ursini2018}. The prominent emission zone originating from the jet also diminishes midway along the total linear extent of the cocoon in this case. Despite the jet's bulk flow experiencing minimal decollimation among the high jet-power runs, its propagation through a rarefied environment results in a mini warm spot formation at the jet-head position (see Section~\ref{Sec:Hotspots and jet dynamics}).

\begin{figure}
\ContinuedFloat
\centering
\includegraphics[width=\columnwidth]{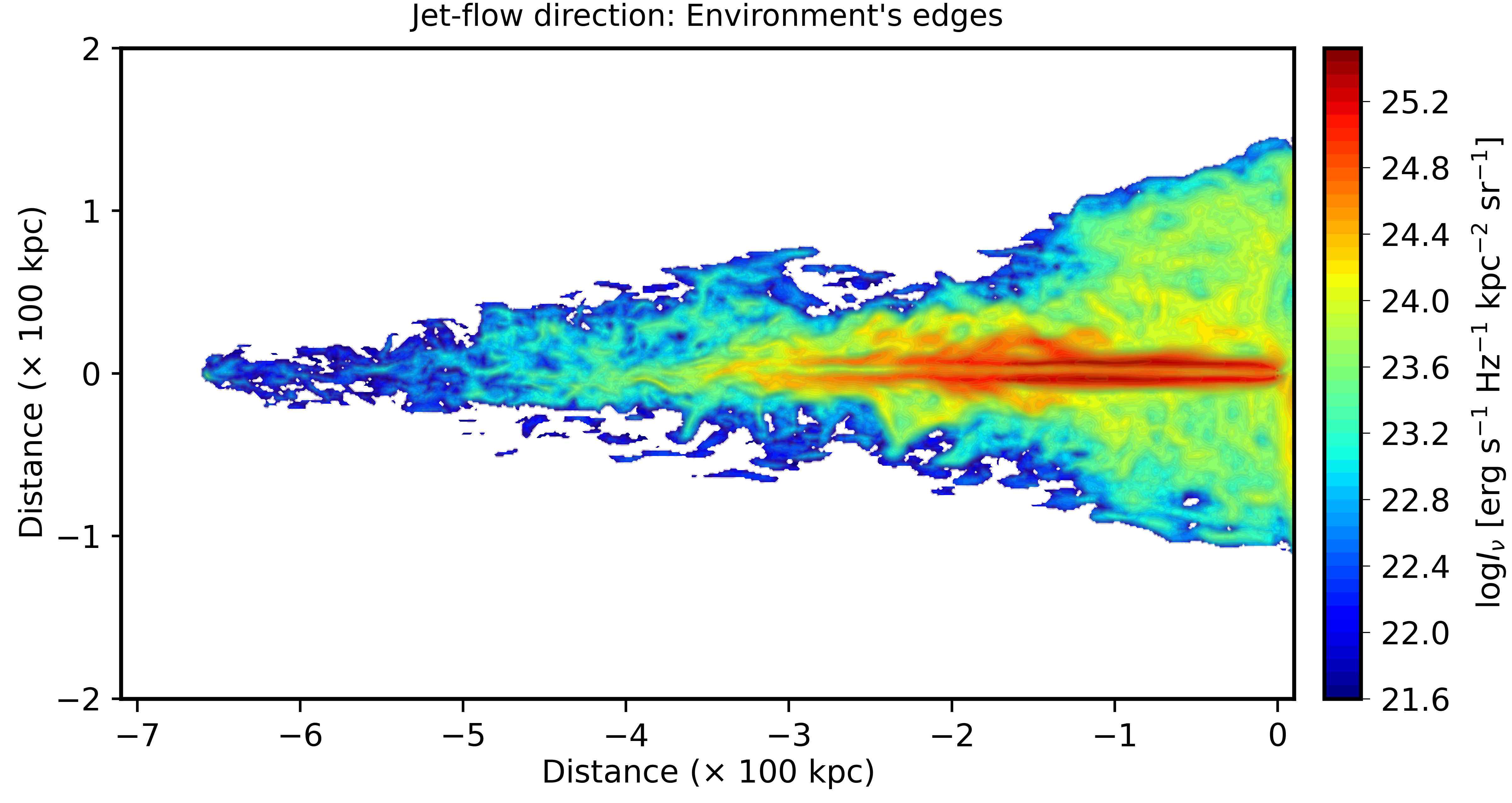}
\caption{C. \textit{Continued.} Intensity map (log $I_{\nu}$) at 1 GHz radio frequency reveals an intriguing case where the jet flows with minimal hindrance due to its propagation along the edges of the surrounding environment ("GRG\_hp\_edge"). The faster propagation, causing increased shear between the cocoon and the environment, as well as between the jet spine and sheath layers, results in the generation of instabilities, leading to the formation of pronounced turbulent structures at 49 Myr.}
\label{Fig:Emission_edge} 
\end{figure}

\subsubsection{Hotspots and jet dynamics}\label{Sec:Hotspots and jet dynamics}
It is worth noting that our emission maps lack prominent hotspots at the jet head in high-power cases. Instead, peak emission appears midway along the jet path, despite sustained thrust to the edge. Hotspots, typically marking the termination shock where powerful jets interact strongly with the ambient medium \citep{Bromberg2011}, form as jets decelerate from near-light-speed at launching sites to $\lesssim 0.1c$ at lobe fronts \citep[for GRGs;][]{Jamrozy2008}. However, their radiative prominence also critically depends on the distribution and strength of the magnetic field \citep{Black1992,Pyrzas2015,Baghel2023}. 

\begin{figure*}
\centering
\includegraphics[width=1.9\columnwidth]{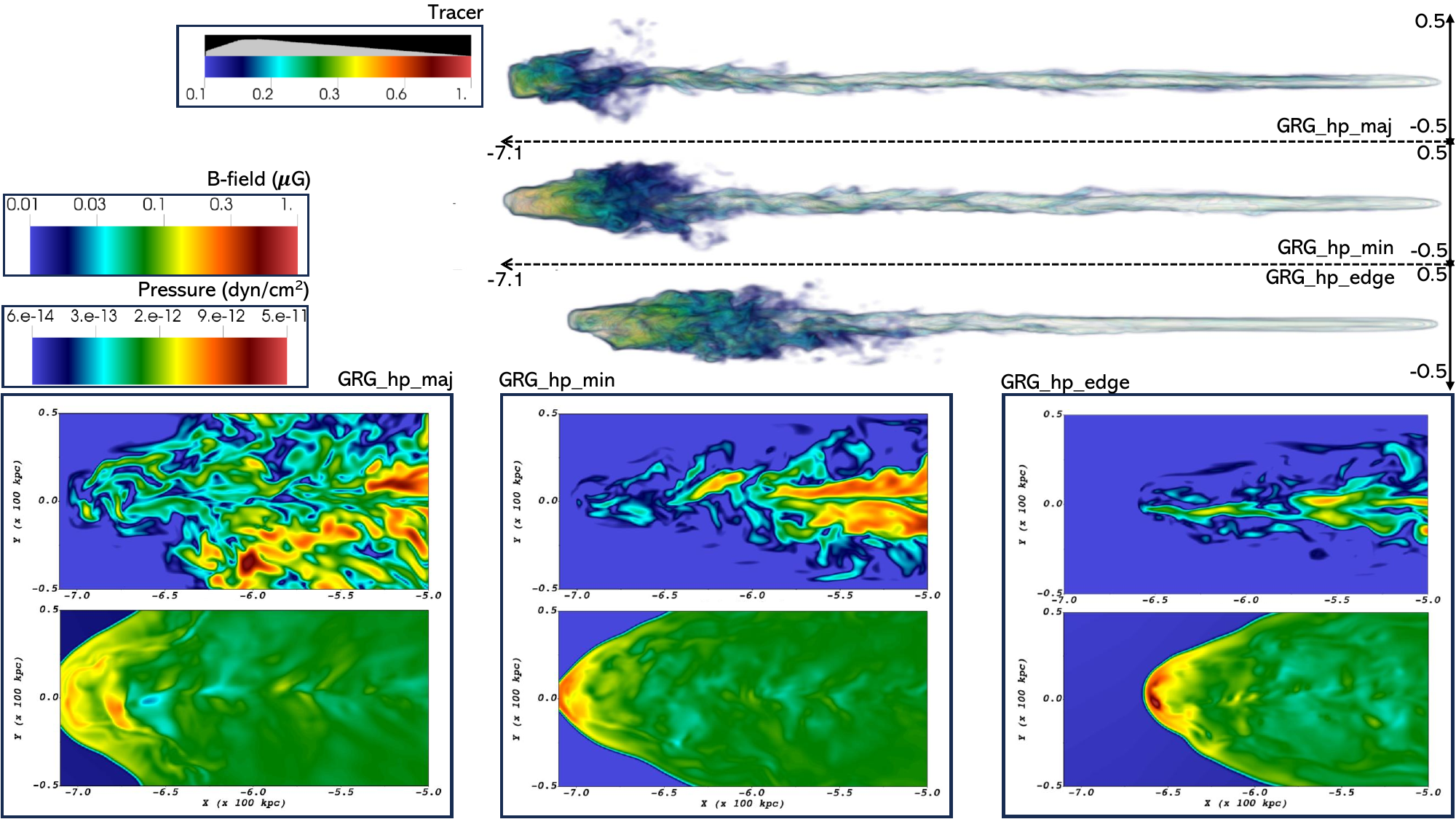}
\caption{Dynamic maps illustrating the jet spine and head structures, elucidating the absence of prominent hotspots in our synthetic emission maps for high-power jet simulations. The \textit{top panel} presents a 3D volume-rendered visualization of the jet tracer, highlighting the jet beam – jet head morphology at their corresponding evolutionary ages. The \textit{bottom panel} shows 2D slices focused on the jet head: the top row maps the magnetic field magnitude, while the bottom row maps the pressure distribution. These maps reveal that although the jet sustains its thrust up to the terminal shock, the magnetic field strength drops sharply at this region—largely advected away by the backflowing plasma—thereby suppressing the formation of bright hotspots.}
\label{Fig:jet_emission} 
\end{figure*}

In Fig.~\ref{Fig:jet_emission}, we present dynamical maps from our high-powered GRG jet simulations. The top panels show the 3D tracer distribution, revealing the evolution of the jet spine and the shape of the head region that primarily drives the backflowing plasma. The bottom panels show 2D slices focusing on the jet head, displaying the B-field magnitude and pressure distribution. Notably, around the midpoint of the jet’s extent, the jet spine begins to lose its collimation and develops mild undulations or wiggling. This induces matter leakage into the surrounding cocoon, generating a spine-sheath morphology beyond mid-distance. Concurrently, the jet head in these high-power cases produces substantial backflowing material.

This continuous leakage and backflow transport B-field lines into the cocoon, which leads to a dilution of the magnetic field strength at the jet head by an order of magnitude or more in comparison with the jet injection zone (weakening the emission)—even while the thrust remains largely intact, as seen in the pressure maps (Fig.~\ref{Fig:jet_emission}). For instance, in the GRG\_hp\_maj case, where the jet experiences greater obstruction by the environment and forms a broader cocoon, the magnetic field becomes more evenly distributed across the cocoon. This explains why its emission map shows lobes illuminated over a wider region near the jet head, accompanied by the formation of a warm spot \citep{Leahy1997} in the top-hat region. By contrast, in the other two high-power runs, the B-field remains concentrated around the jet sheath, leading to emission that is largely confined along the jet beam and gradually fades as it approaches the jet head. These results point to the possibility that some fraction of GRGs are likely larger than current observations indicate. As radio surveys advance in sensitivity and resolution, they could expose the full extents of these sources and identify additional GRGs now misclassified as smaller radio galaxies \citep{Hardcastle2019,Koribalski2025}. 

To further assess the weakening of magnetic field strength along the jet, particularly at the terminal regions, we computed the plasma beta ($\beta = 8\pi P_{\text{jet}}/B^2_{\text{jet}}$) in the vicinity of the jet head. Our analysis reveals $\beta \gtrsim 20$ in this region, indicating that the magnetic pressure plays a subdominant role compared to the thermal pressure at the jet terminus. Notably, in the cocoon enveloping the head, $\beta$ rises even further—reaching values $\sim 500$ or higher—suggesting significant magnetic field dilution. This trend, along with the increasingly tangled nature of the magnetic field topology, is further illustrated in Fig.\ref{Fig:Dynamic-Bmap} in Section~\ref{Sec:Polarization measures at 1 GHz}. 

Such features recovered in our GRGs — including the emergence of top-hat–shaped warm spots, and peak emission zones forming well within the lobe boundaries — have also been documented in several other observed GRG cases, such as: \citet{Clarke2017,Chen2018,Sebastian2018,Oei2024_7mpc}. However, to better understand the typical occurrence of hotspots at the jet termini in powerful GRGs, further numerical investigations are needed \citep[e.g., through high-resolution simulations of jet flows capable of modelling flow bifurcation, branching, and complex spine–sheath dynamics;][]{Laing2015,Horton2023}. As suggested in \citetalias{Giri2025}, a comprehensive understanding of these complex emission features requires systematic exploration across a broader parameter space—such as higher jet-to-ambient density contrasts ($\rho_j/\rho_0$) and stronger injected magnetic fields—awaiting future study. In this context, we note that we have initiated a related investigation focused on the early evolutionary stages of GRGs (evolving on the galactic scales), examining the role of magnetic field strength in jet collimation and consequent hotspot formation (Giri et al., in prep.). Additionally, the adoption of advanced emission mapping techniques, such as a Lagrangian-Eulerian particle-fluid framework on high-resolution grids, is planned in future work, facilitated by recent GPU-accelerated code developments.



\subsection{Quantitative exploration of the emission maps}\label{Sec: Quantitative exploration of the emission maps}

\subsubsection{Insights from the lateral intensity variation}\label{Sec:Insights from the lateral intensity variation}
\begin{figure*}
\centering
\includegraphics[width=2\columnwidth]{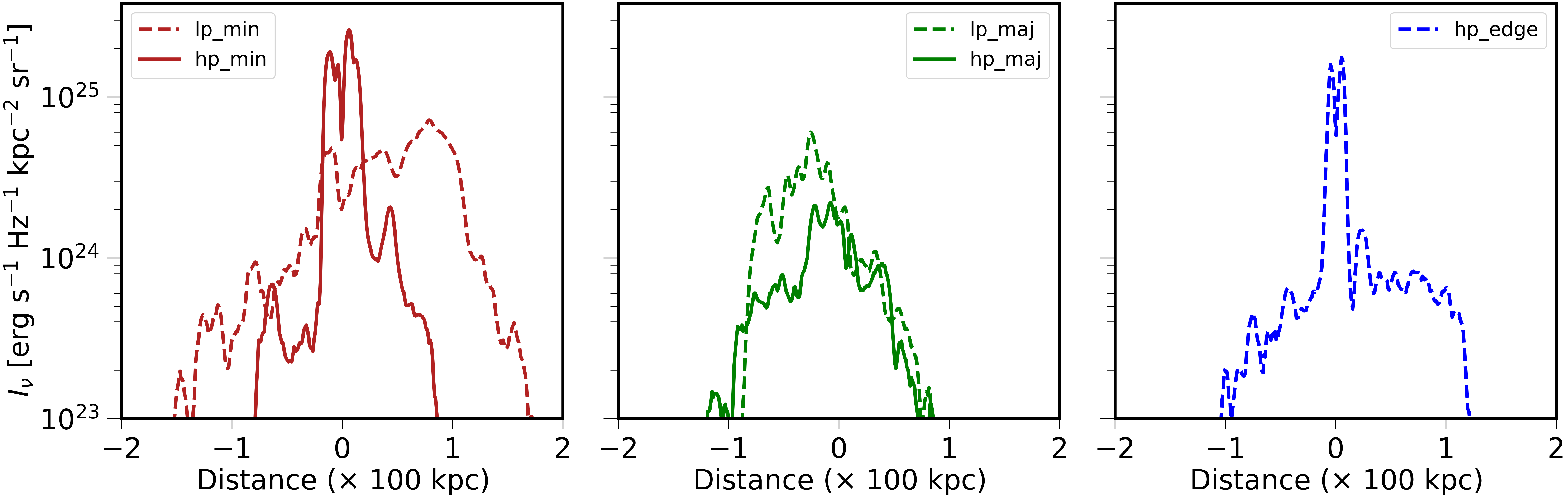}
\caption{Lateral intensity variation measured at the maximum cocoon width for five simulation scenarios (Fig.~\ref{Fig:Emission_minor}(A$-$C)), providing a quantitative assessment of intensity distribution and identifying active zones within the cocoon. Subplots correspond to different simulation runs categorized by `min', `maj', and `edge' cases, enabling a comparison of jet power on cocoon activity and evolution.}
\label{Fig:Sliced_emission} 
\end{figure*}

We begin with the intensity profiles, as shown in Fig.~\ref{Fig:Sliced_emission}, illustrating the lateral emission variation of the simulated structures at their evolved stages. The variation is assessed at the maximum cocoon width, representing a general demonstration of the underlying radio galaxy \citep{Subrahmanyan1996,Malarecki2013,Hardcastle2018}. To ensure a fair comparison within the cocoon and exclude the intensity variations around its immediate edges, we apply a threshold of $10^{23}$ erg s$^{-1}$ Hz$^{-1}$ kpc$^{-2}$ sr$^{-1}$ to the intensity values. In all cases, the intensity values are notably higher around the jet-spine structure, reflecting its active flow and thrust. A shift in the peak emission position from the central axis or mid-point is observed in low jet power scenarios, attributed to the bending of the jets, as can be seen in the dynamical maps shown in Fig.~\ref{Fig:Velocity_dynamics}. 

The most notable feature in this lateral intensity analysis (Fig.~\ref{Fig:Sliced_emission}) is the intensity variation observed in regions extending outward from the jet spine. For both high and low jet power cases, the emission decreases sharply from a relatively high value to the threshold intensity as one moves laterally from the jet spine \citep{Condon2021}. This suggests that as long as the jet remains active, the cocoon structure retains its dominance across a substantial lateral extent of the radio galaxy. Conversely, as the radio galaxy slows in its evolution (e.g., through adiabatic expansion or cessation of jet activity) or undergoes significant entrainment, the emission gradually diminishes outward from the primary flow, as suggested by the observational findings of \citet{Subrahmanyan2008}. This gradual reduction in lateral emission may therefore serve as an indicator of intermittent activity or lower active stages of radio galaxies, complementing other diagnostic parameters such as core dominance or hotspot position estimation \citep[e.g.,][]{Subrahmanyan2008,Cotton2020}.

\subsubsection{Insights into the radio power–jet power correlation}
\begin{figure}
\centering
\includegraphics[width=\columnwidth]{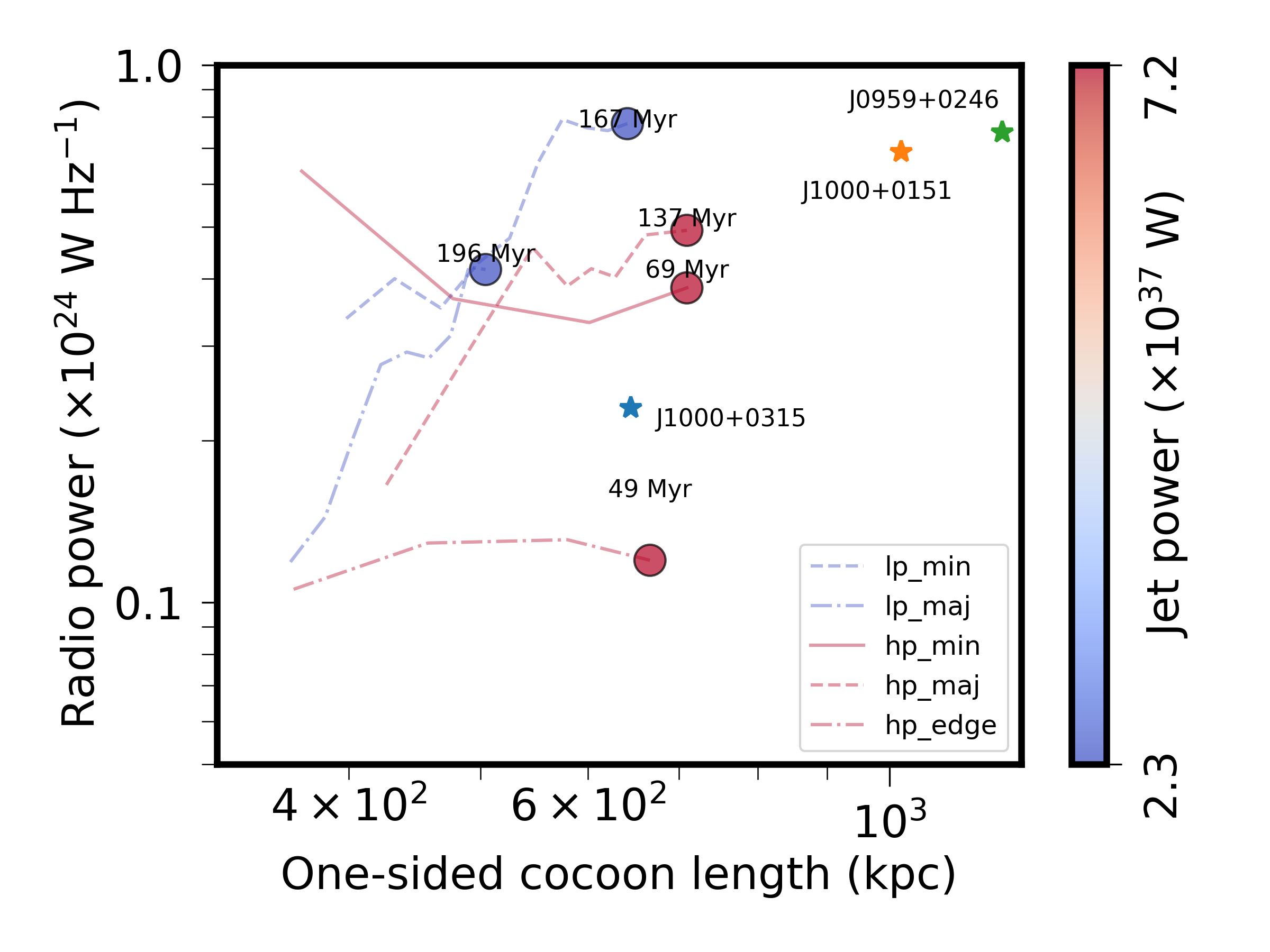}
\caption{Radio power at 1 GHz of the (one-sided) cocoon versus (one-sided) cocoon length, presented at the evolved stages (dynamical ages attached) of the simulated GRGs. Two jet power values used in the simulations are also highlighted, in blue and red. The evolutionary track of the simulated GRGs in the Power-Distance (\textit{P-D}) diagram is also shown from the point they exceed 350 kpc, qualifying as GRGs in their total extent. For reference, we include measurements of three observed GRGs from \citet{Charlton2024} in similar observing band, representing a typical one-half estimation. We note that the above calculation does not explicitly account for the cooling effects.}
\label{Fig:R-J_Power} 
\end{figure}

Motivated by the growing interest in the radio power versus linear size (\textit{P-D}) distribution for GRGs \citep{Andernach2021,Simonte2022}, we analyzed this trend  (see, Fig.~\ref{Fig:R-J_Power}) using our five simulated outputs at their evolved stages (one-sided length exceeding 350 kpc). Considering the evolution track of each simulations, for the lower jet power scenarios (depicted in blue), we observe an increase in radio power with increasing source age. The high jet power situations (depicted in red), however, exhibit a variable trend, with total radio power sometimes increasing or decreasing with age. A closer examination suggests that this behavior correlates with the extent and shape of the radio galaxy, i.e., sources with wider and more extended structures, also indicative of prominent lobe formation, tend to exhibit higher radio power for a given jet power. This trend may not be universally applicable to all GRGs \citep{Ishwara-Chandra1999}, as the influence of radiative cooling becomes increasingly significant with the source's growing size and redshift evolution—an effect not explicitly accounted for in our calculations. In the following sections, we briefly examine the relative dominance of various cooling mechanisms in shaping the radiative properties of GRGs.

Our observation of Fig.~\ref{Fig:R-J_Power} also suggests that sources with greater linear extents do not exhibit any trend in radio power values, with some showing comparable or even lower radio power estimates \citep[e.g.,][]{Dabhade2020_SaganI,Simonte2024}. This observation hints that the combination of radio power and linear extent alone is insufficient to fully capture the complexities of processes within a GRG system. It underscores the importance of considering additional metrics, such as the relationship between radio power and the projected emission area of the cocoon (\textit{P-A}), to gain deeper insights (also reiterating from the above paragraph). 

The \textit{P-D} analysis (Fig.~\ref{Fig:R-J_Power}) also emphasizes a key conclusion: using a simplistic analytical formulation to relate radio power and jet power \citep[e.g.,][]{Hardcastle2018} for statistical GRG samples \citep{Dabhade2020_SaganI} may lead to inaccurate estimates. This is due to the formulation's dependence on various factors, including the complex geometry and associated parameters of the ambient environment, which significantly influence the jet-induced morphology and radio power \citep{Mingo2019}. These factors, which vary depending on the parametric space where the GRG formed, make it unsuitable for straightforward application to individual GRGs.

The radio power values obtained at 1 GHz for the simulated GRGs are slightly on the lower side. Adopting a redshift of 0.05, the estimated flux values for these simulations are $\sim$ 135 mJy ("lp\_min"), 67 mJy ("hp\_min"), 73 mJy ("lp\_maj"), 86 mJy ("hp\_maj") and 21 mJy ("hp\_edge"), respectively. It is important to note that these values correspond to the one-sided cocoon only, and thus, the total values are likely to be higher. Additionally, we note that the emission maps were generated under the assumption that 10\% of the internal energy is converted into emission (Section~\ref{Sec:Post-processing I: Generating intensity maps}). The mentioned ranges of the estimated radio power and flux values, however, are relevant with those reported for several GRGs in similar frequency bands, as referenced in, \citet{Machalski2009,Machalski2011,Andernach2021,Delhaize2021,Charlton2024}. For reference, in Fig~\ref{Fig:R-J_Power}, a representative one-sided radio power and length estimation of three GRGs from \citet{Charlton2024} is shown.

\subsubsection{Magnetic field estimates and their implications}\label{Sec:Magnetic field estimates and their implications}
\begin{figure}
\centering
\includegraphics[width=\columnwidth]{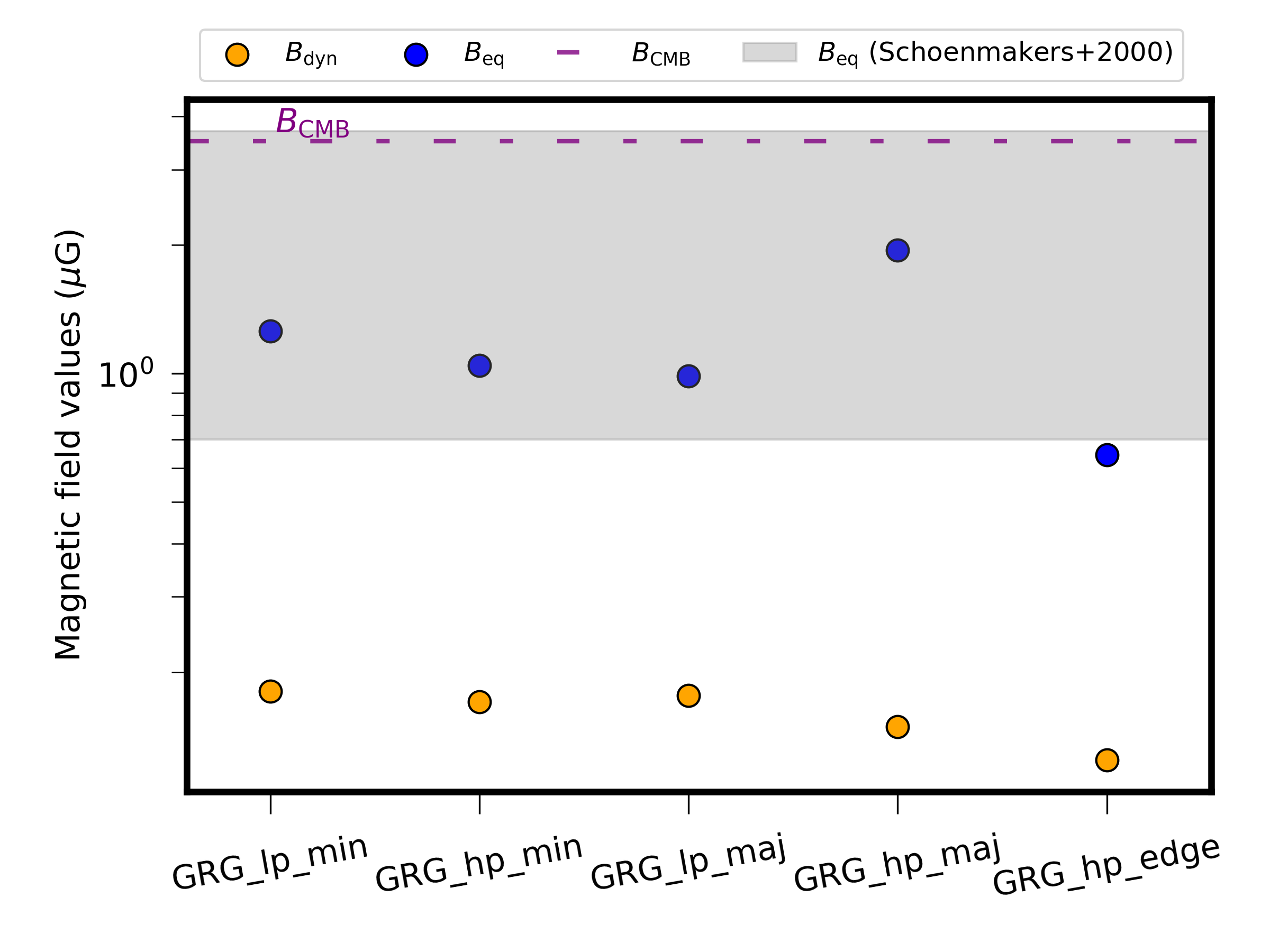}
\caption{Estimation of the strengths of various magnetic field values (in $\mu$G) for our five simulated scenarios, as shown on the horizontal axis. The magnetic fields included are the dynamical or intrinsic simulated B-field ($B_{\rm dyn}$), the equipartition magnetic field ($B_{\rm eq}$), and the estimated CMB-equivalent magnetic field strength ($B_{\rm CMB}$) at a redshift of 0.05. The gray zone highlights the range of $B_{\rm eq}$ values adopted for a sample of GRGs; details in Section~\ref{Sec:Magnetic field estimates and their implications}}.
\label{Fig:B_comparison} 
\end{figure}

To infer magnetic field strengths critical for estimating GRG parameters like age, we calculated and compared different B-field estimates, as shown in Fig.~\ref{Fig:B_comparison}. The first estimate, shown as purple points, represents the median values of dynamically simulated magnetic fields ($B_{\rm dyn}$) at the evolved times of the five simulations, observed to vary around 0.15 $\mu$G (see also, \citetalias{Giri2025}). 

The second estimate represents the median magnetic field values calculated using the equipartition assumption ($B_{\rm eq}$), which assumes equal energy distribution between radiating non-thermal particles and the magnetic field. Following \citet{Hardcastle2002}, we deduce the $B_{\rm eq}$ values as follows,
\begin{equation}
    m_e c^2 \int_{\gamma_{\rm min}}^{\gamma_{\rm max}} \gamma \, N(\gamma) \, d\gamma = \frac{B^2_{\rm eq}}{8\pi}
\end{equation}
where $N(\gamma)$ is the non-thermal electron distribution, with energy represented by the Lorentz factor $\gamma$, ranging from $10^2$ to $10^6$, as discussed in Section~\ref{Sec:Post-processing I: Generating intensity maps}. The equipartition field strengths range from 0.65 $\mu$G to 1.94 $\mu$G \citep{Schoenmakers2000,Andernach2021,Machalski2008}, with the ratio of dynamical to equipartition values as 0.15 ("lp\_min"), 0.17 ("hp\_min"), 0.18 ("lp\_maj"), 0.08 ("hp\_maj"), and 0.2 ("hp\_edge") \citep[e.g., see the conclusion on powerful radio galaxies;][]{Mahatma2020}. For reference, we highlighted the $B_{\rm eq}$ values for a sample of GRGs from \citet{Schoenmakers2000} in a gray zone in Fig.~\ref{Fig:B_comparison} \citep[see also,][]{Tamhane2015,Charlton2024}, illustrating their similarity to our estimated values, while the true intrinsic values are nearly an order of magnitude lower. The review by \citet{Dabhade2023} summarizes a range of \( B_{\rm eq} \) values from 1 to 16 $\mu$G, with a median of 5 $\mu$G. However, when compared to X-ray measurements (independent of the equipartition assumption), this appears to be nearly an order of magnitude lower, further emphasizing our reported situation.

With the assumption that the spectral age follows the relation, t$_{\rm spec} \propto B^{-3/2}$ \citep{VanDerLaan1969}, the differences in equipartition values, which vary by nearly an order of magnitude from the dynamical values, can lead to significant discrepancies in the spectral age compared to the true dynamical age (with the spectral age being smaller). Thus, the inherent (dynamical) age of many GRGs may need to be re-evaluated, which might highlight discrepancies among various age estimation methods \citep[see, e.g.,][]{Machalski2009,Machalski2011}. This re-assessment is crucial for testing the proposed formation mechanisms of GRGs under the standard jet evolution hypothesis and clarifying the inconsistencies observed across studies concerning the GRG ages, such as \citet{Jamrozy2008,Komberg2009,Sebastian2018,Marecki2021}.

The third estimate is the CMB-equivalent magnetic field strength ($B_{\rm CMB}$) at a redshift ($z$) of 0.05, calculated using the formula,
\begin{equation}
    B_{\rm CMB} = 3.18 \times (1 + z)^2 \, \, \mu {\rm G}
\end{equation}
The equivalent level is shown with a dashed line in Fig.~\ref{Fig:B_comparison}, indicating higher values than both \( B_{\rm dyn} \) and \( B_{\rm eq} \), suggesting the dominant role of the inverse-Compton cooling process in GRGs, which remains significant even at this low-redshift regime \citep[aligning with, e.g.,][]{Ishwara-Chandra1999,Schoenmakers2000,Subrahmanyan2008}.

\subsubsection{Influence of matter entrainment on morphological appearance} 
The entrainment of ambient matter into the cocoon primarily arises from the significant velocity shear and density contrast between the relatively static ambient medium and the dynamic jet-cocoon system. This process is facilitated by the growth of instabilities, including Kelvin-Helmholtz \citep{Borse2021} and Rayleigh-Taylor instabilities \citep{Abolmasov2023}, combined with the effects of large-scale cocoon rotation \citep{Gourgouliatos2018}. These mechanisms drive turbulent mixing, enabling the transport (beside momentum exchange) of ambient material deep into the cocoon and into the vicinity of the jet-beam structure, significantly impacting the radiative properties of the system.

\begin{figure}
\centering
\includegraphics[width=\columnwidth]{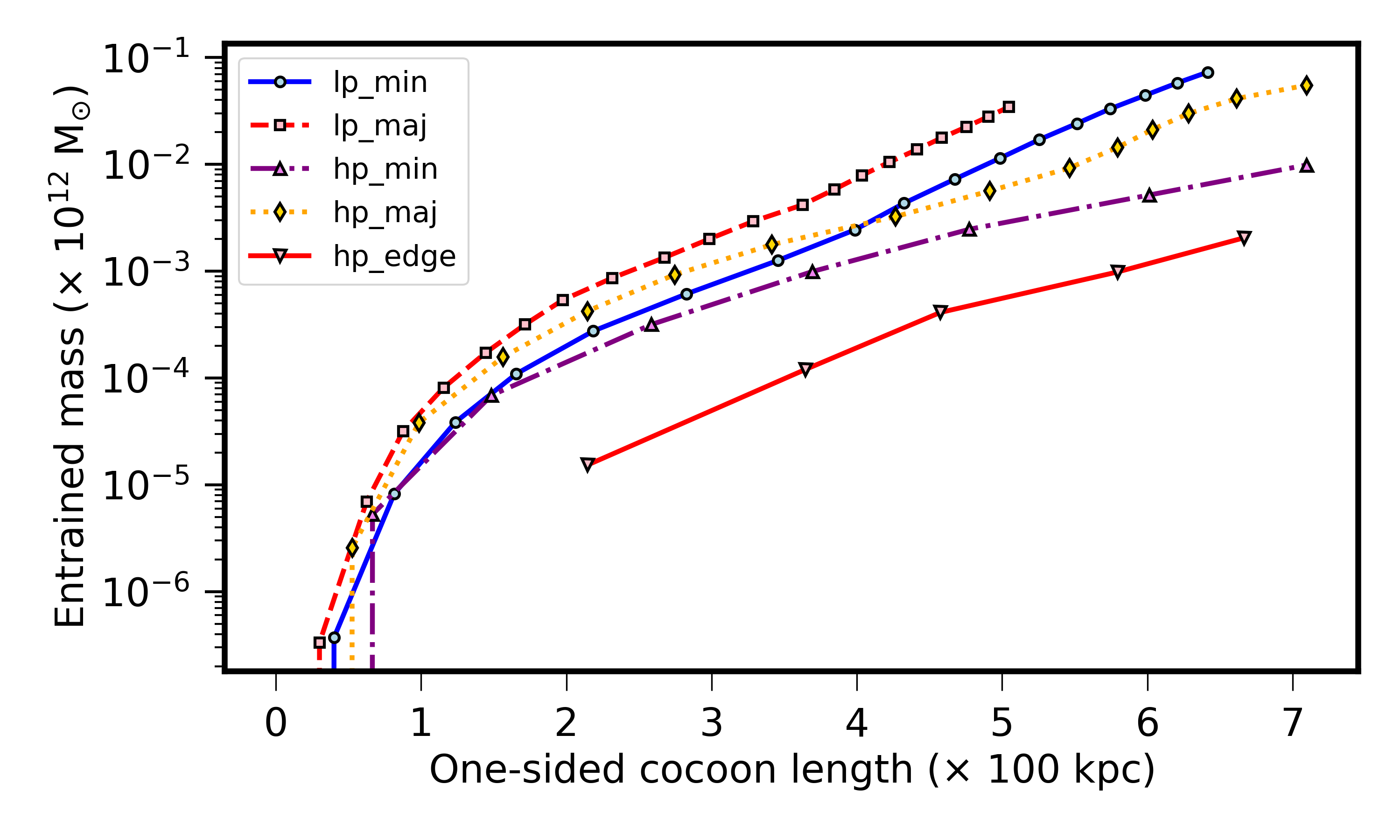}
\caption{Amount of ambient medium matter entrained into the cocoon for the five simulated scenarios is presented, with the evolution plotted against the linear growth of the cocoon (also mimicking the temporal growth).}
\label{Fig:me} 
\end{figure}

In our simulated systems, we estimated the amount of total entrained ambient matter using the tracer $Q_1$ (utilising the location and corresponding $Q_1$ fraction) within the boundary of the jet-induced cocoon, defined by $Q_2 \geq 10^{-7}$ \citep{Mukherjee2020}, as shown in Fig.~\ref{Fig:me}. We reiterate that $Q_1$ and $Q_2$ are passive scalars used to trace different media: $Q_1$ for ambient gas and $Q_2$ for jet gas (see Section~\ref{Sec:Revisiting Paper I}). The results reveal an initial rapid phase of entrainment, attributed to the jet's propagation through the denser core region, followed by a steady increase in the entrained mass as the cocoon length increases. The encapsulated mass represents a considerable fraction of the underlying ambient medium (baryonic mass of the ambient medium $\sim 3.3 \times 10^{12} M_{\odot}$) at the evolved times. 

In the low-powered scenarios (Fig.~\ref{Fig:me}), the jet decollimation and wider lobe formation results in higher entrainment. This behavior is expected and is suspected to be a key factor in the jet's rapid deceleration, leading to its transition from a collimated jet flow to an FR-I-like lobe morphology \citep{Massaglia2016,Abolmasov2023,Rossi2024}. Similarly, in the high jet-power case with increased jet opposition ("GRG\_hp\_maj"), where the formation of a mini-lobe structure is observed in the frontal part of the flow, the accumulation of ambient matter remains comparable. The other high-powered situations show lower levels of matter accumulation, as the high-speed flow helps maintain the collimation of the jet spine for a longer duration. However, due to the increased velocity shear between the ambient medium and the cocoon, and also being significantly less dense (with the injected jet density being $10^{-5}$ times that of the ambient density), the cocoon begins to experience disruption, although the central spine remains relatively protected \citep{Rossi2008}. A direct consequence of this phenomenon is illustrated in Fig.~\ref{Fig:Emission_minor} (A: lower-panel), which shows the described jet-spine and cocoon behavior. The lowest mass entrainment occurs when the jet flows along the edges of the ambient medium (Fig.~\ref{Fig:me}), resulting from its propagation through a lower-density region and the collimation of the flow due to rapid propagation. However, following the processes noted earlier, cocoon disruption is evident in this case also.

Interestingly, for the "GRG\_hp\_edge" case (Fig.~\ref{Fig:Emission_edge}(C)), the jet-spine is observed to leak matter into the cocoon-sheath structure (around $\sim 200$ kpc), offering insight into the turbulent mixing of non-thermal, older back-flowing particles with freshly injected ones. The enhanced shear between the jet's spine-sheath structure, along with instabilities from recollimation shocks, likely contributes to these features. While understanding these phenomena is crucial for spectral studies of radio galaxies \citep{Turner2018}, we have not focused on them in detail due to the need for significantly higher resolution for convergence, warranting a separate study. Previous works have highlighted the importance of such investigations in greater detail \citep{Borse2021,Abolmasov2023,Wang2023}, which should be extended to GRGs, as jet stability is vital for their survival over large scales \citep[since greater matter entrainment leads to enhanced jet deceleration;][]{Rossi2024}.

\subsection{Correlating emission features with dynamical properties: Filamentary extensions} \label{Sec:Correlating emission features with dynamical properties: Hotspot formation and filamentary extensions}

Observations of extragalactic radio jets with contemporary telescopes have unveiled novel features in a growing number of sources, broadly categorized as `filamentary threads', whose origins remain a topic of ongoing debate \citep{Rudnick2022}. Prominent and extended filaments are often observed emanating from the cocoon structures of these jets, extending outward into the surrounding intracluster medium where the radio source is embedded \citep{Ramatsoku2020,Rudnick2022,Koribalski2024}. 
In addition, application of advanced edge-detection techniques have revealed a network of intricate filamentary structures within the cocoon as well \citep{Condon2021,Velovic2023}. These internal filaments exhibit complex geometries, often intertwining or displaying irregular patterns, suggesting the interplay of various dynamical processes, such as turbulence, shear flows, and magnetic field compressions \citep{Rudnick2021,Upreti2024,Wezgoweiec2024}. 

Recently, observations of two GRGs with the MeerKAT radio telescope\footnote{MeerKAT telescope: \url{https://www.sarao.ac.za/science/meerkat/}} have unveiled intricate internal \citep{Cotton2020} and extended external filamentary structures \citep{Cotton2025}, emphasizing the need for a deeper investigation into their properties and origins in such large-scale jetted sources. These structures may offer critical insights into questions such as their potential role in non-thermal particle transport and re-acceleration. In this context, we have sought to identify correlations between the filamentary structures observed in our formulated emission maps (Fig.~\ref{Fig:Emission_minor}(A$-$C)) and the underlying three-dimensional dynamics responsible for their formation.

The emission maps for the five simulated cases$^{\ref{fn:Sim_Cases}}$ were analyzed for this purpose. A detailed examination of these maps (Fig.~\ref{Fig:Emission_minor}(A$-$C)) summarized that the bulk flow of the jet-beam undergoes substantial widening, particularly with respect to its injection radius (i.e., 1 kpc). This behavior is especially prominent in the low jet-power cases and occurs after the jet propagates a certain distance within the cocoon, well before reaching its full extent. This widening gives rise to filamentary extensions with different geometries, including tendril-shaped or finger-like structures, caldera-like formations characterized by emission valleys surrounded by enhanced emission ridges, and large vortex ring-like features spanning the lobe.

\begin{figure*}
\centering
\includegraphics[width=2\columnwidth]{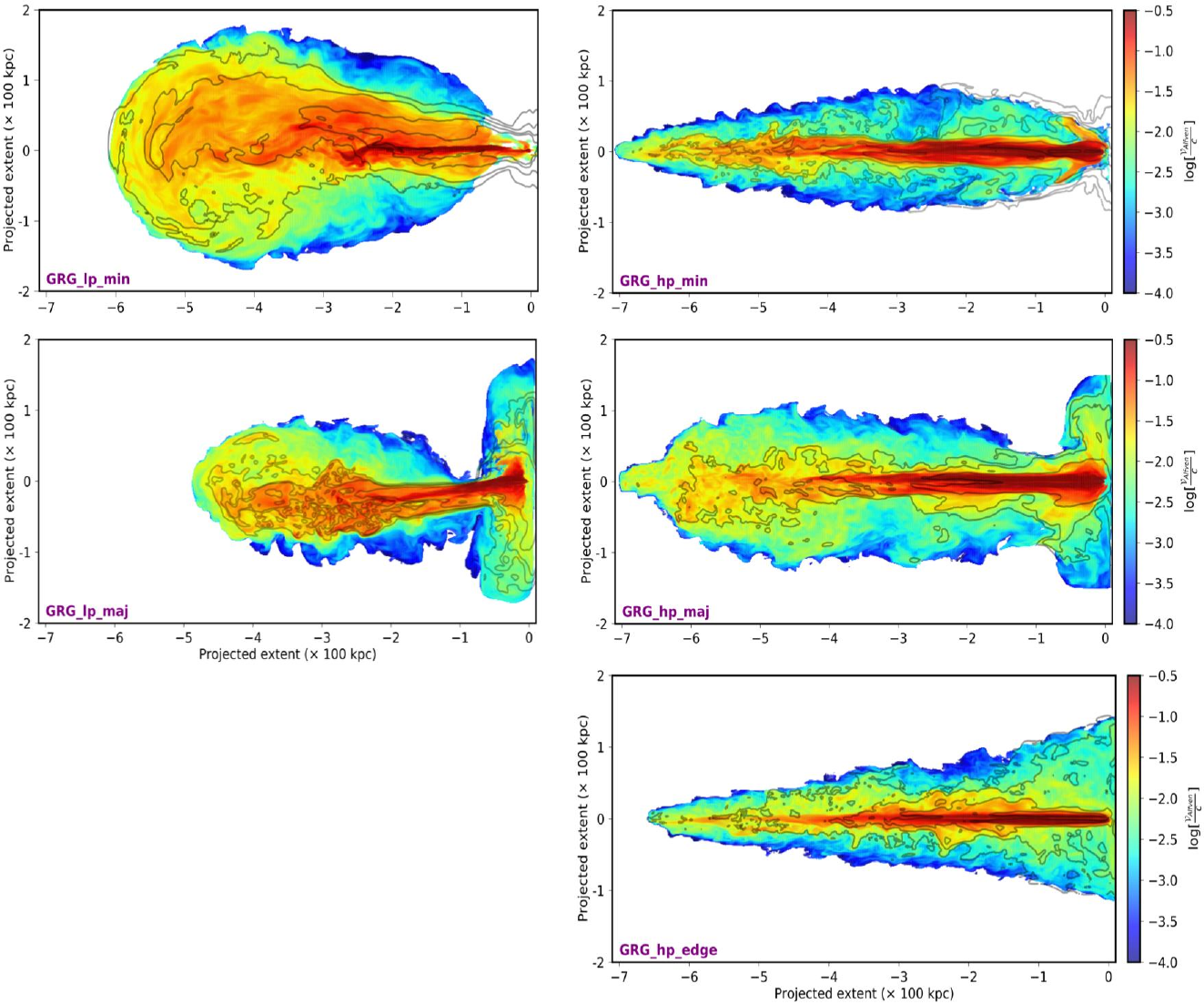}
\caption{Projected colormaps depict the distribution of Alfvén velocity ($v_\text{Alfv\'{e}n}$; normalized by light-speed) on a logarithmic scale, overlaid with emission contours derived from the maps presented in Fig.~\ref{Fig:Emission_minor}(A$-$C). These combined maps enable the identification of a strong correlation between dominant emission features, such as the emission filaments, and their probable dynamical origins (e.g., the magnetized flow) across the five simulation cases, as indicated in the individual sub-figures. The 3D maps of Alfvén speed is projected onto the $x-y$ plane to allow direct comparison with the emission maps with their line-of-sight observation aligned along the $z-$axis.}
\label{Fig:Alfvenic_Velocity} 
\end{figure*}

Despite such complex geometric appearances observed in the projected 2D maps, we found striking similarities to these emission layers in their three-dimensional counterparts data, when analyzing the variation of the Alfvénic speed within the jet-cocoon geometry. The Alfvén speed ($v_\text{Alfv\'{e}n}$) is a key indicator of magnetized flow, defined as follows (given the sub-relativistic evolution of the cocoon material; \citetalias{Giri2025}),
\begin{equation}
    v_\text{Alfv\'{e}n} = \frac{B}{\sqrt{4\, \pi\, \rho}}
\end{equation}
where $B$ and $\rho$ represent the magnetic field strength and the plasma density at each location within the cocoon, respectively. To ensure a fair comparison with the line-of-sight integrated emission map (where the line of sight is along the $z-$axis), we project the 3D distribution of $v_\text{Alfv\'{e}n}$ onto the $x-y$ plane (by averaging and collapsing all values along the $z$-direction onto the $z=0$ plane), as the dominant 3D layers of the jet-cocoon system will also reflect themselves in the projected map. This is shown in Fig.~\ref{Fig:Alfvenic_Velocity}, where the projected $v_\text{Alfv\'{e}n}$ maps were subsequently overlaid with emission contours, with the contour levels carefully selected for each case to prominently highlight the filamentary structures noticed in the intensity maps (Fig.~\ref{Fig:Emission_minor}(A$-$C)). For instance, in the "GRG\_lp\_min" case, the contour levels in log($I_{\nu}$) (in units of erg s$^{-1}$ Hz$^{-1}$ kpc$^{-2}$ sr$^{-1}$) were set to [23.8, 24.45, 25.0, 25.25], while for the "GRG\_lp\_maj" case, the levels were chosen as [24.08, 24.45, 24.6, 24.9].

The striking resemblance between the projected $v_\text{Alfv\'{e}n}$ maps and the emission contours, which trace the filamentary geometries, reveals two critical insights. First, the widening of the jet into the broader cocoon flow appears to be the primary driver of filament formation, with jet instabilities playing a comparatively minor role. Second, these filaments correspond to magnetized flows, effectively acting as magnetic threads within the cocoon. The first finding underscores the importance of turbulence-driven particle transport and acceleration within the cocoon, suggesting that this process could be a significant contributor to the observed radio emission in giant radio galaxies \citep[e.g., see,][for relevance in smaller radio galaxies]{Mukherjee2021,Kundu2022}. The second insight is particularly compelling, as it provides, to our knowledge, one of the most recent numerical indications of the magnetized nature of filaments \citep[consistent with the observed B-field pattern of external filaments;][]{Rudnick2022,Velovic2023}. The magnetization of these filaments, which is anticipated to manifest in polarization maps, has been observed and is discussed in detail in Section~\ref{Sec:Polarization measures at 1 GHz}.

\begin{figure*}
\centering
\includegraphics[width=2\columnwidth]{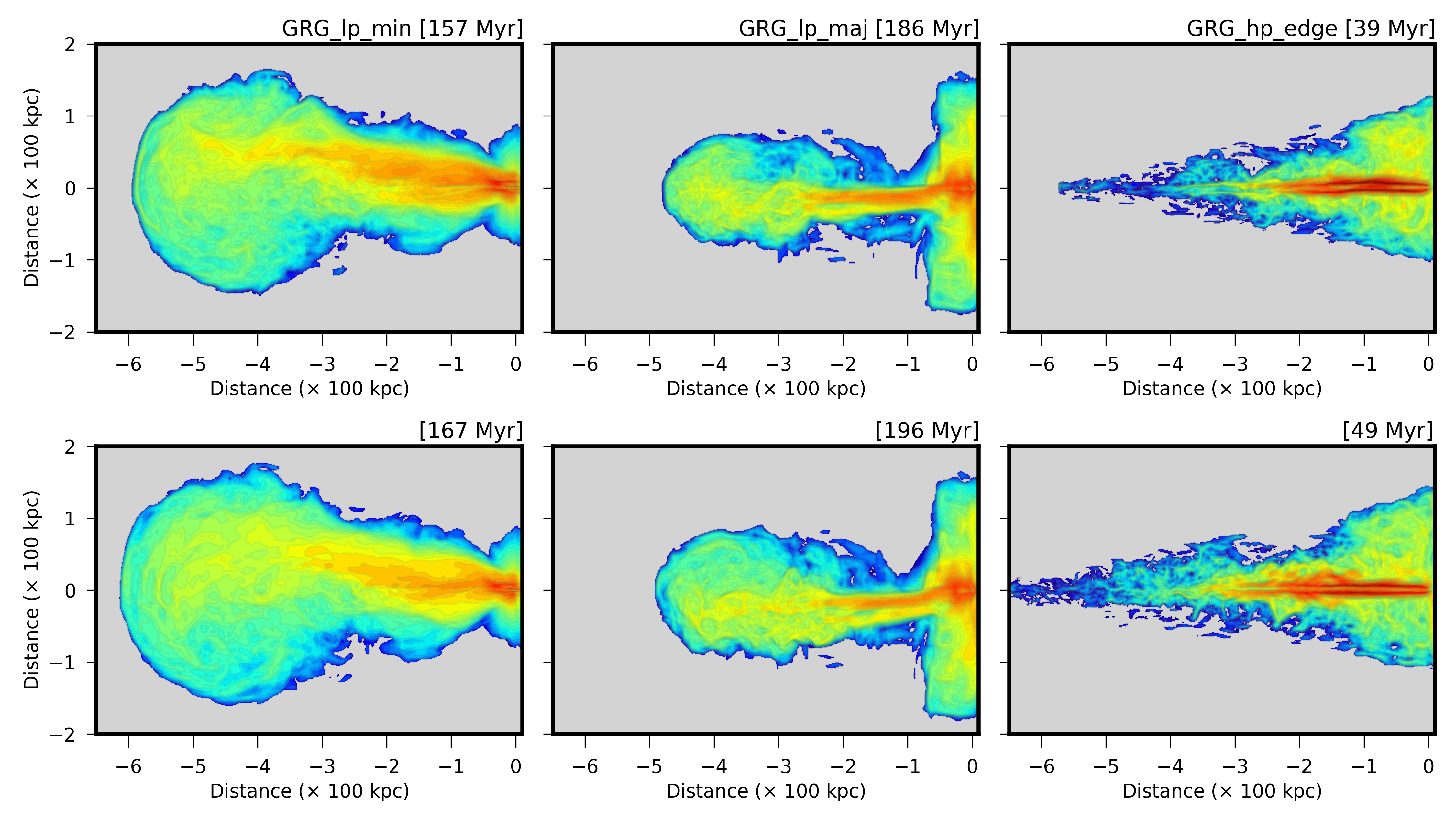}
\caption{Emission maps for three simulations (labeled in the sub-figures) at times $\sim$10 Myr earlier than those presented in Fig.~\ref{Fig:Emission_minor}(A-C), reiterated in the bottom row. A comparison between these temporal snapshots underscores the dynamic and transient nature of filamentary structures within the cocoon, while also aiding in constraining the typical lifetimes of these multi-scaled internal features. For further details, refer to Section~\ref{Sec:Correlating emission features with dynamical properties: Hotspot formation and filamentary extensions}.}
\label{Fig:Varying_filaments} 
\end{figure*}

We investigated the typical lifetimes of these filaments within the cocoon and found them to be highly dynamic, with their geometry and appearance evolving on timescales significantly shorter than the lifetime of the GRG age. In Fig.~\ref{Fig:Varying_filaments}, we present intensity maps from three simulations at an earlier epoch, $\sim$10 Myr before (\textit{top-row}) the snapshots shown in Fig.~\ref{Fig:Emission_minor}(A-C) (\textit{bottom-row}), corresponding to the data-saving interval of our simulations. The comparison between the two snapshots demonstrates that the most prominent filaments within the cocoon, formed directly by the channeling of the jet's bulk flow, undergo significant topological changes on timescales shorter than 10 Myr. To constrain this variability further, we conducted a higher temporal-resolution simulation with data saved every 0.3 Myr intervals for "GRG\_hp\_min"$^{\ref{fn:Sim_Cases}}$ (a computationally intensive task, hence performed for one case only). Analysis of these better time-resolved emission maps reveals that prominent large-scale filamentary structures (distinctly identifiable through visual inspection) by their formation via jet flow branching and direct matter channeling, have estimated lifetimes of approximately 3–10 Myr. Intermediate-scale filaments, arising from jet's bulk flow widening and decollimation, are comparatively shorter-lived, with lifetimes of $\sim$0.9–3 Myr. In contrast, smaller-scale filaments, identified qualitatively and less prominent in nature, are predominantly located farther from the bulk jet beam, contributing to the dynamic cocoon formation, which are highly transient, with lifetimes shorter than 0.3 Myr. To better constrain the survival age of these transient structures, higher-resolution data and significantly faster data-saving intervals are required.

\section{Polarization measures at 1 GHz}\label{Sec:Polarization measures at 1 GHz}

\begin{figure*}
\centering
\includegraphics[width=2.05\columnwidth]{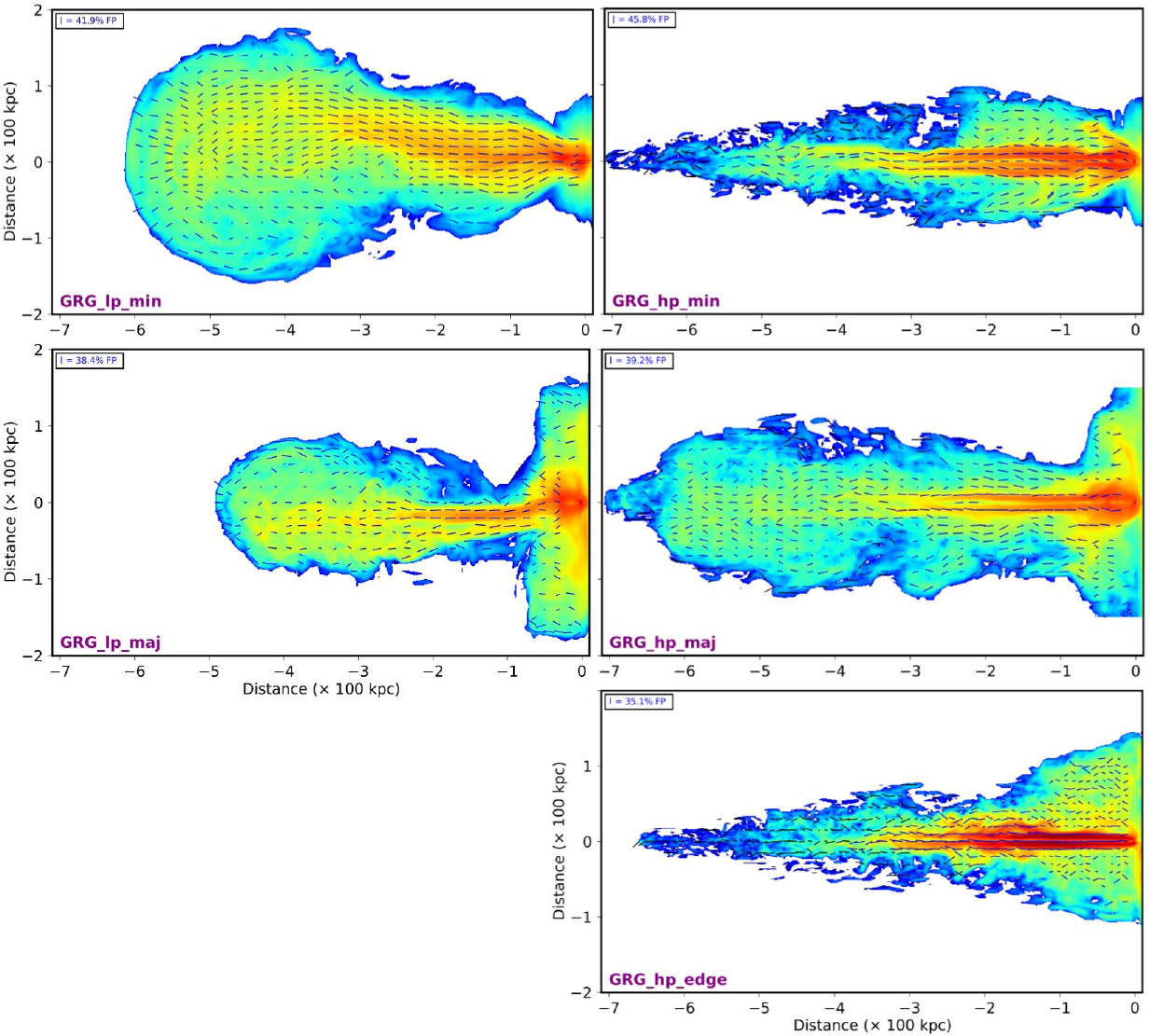}
\caption{Intensity variation maps (as background) for five simulated scenarios at their evolved stages, presented as colormaps, overlaid with projected magnetic field distributions. The line lengths indicate proportionality to the corresponding fractional polarization (FP) values at each point. Median FP values are noted for `GRG\_lp\_min' (41.9\%), `GRG\_hp\_min' (45.8\%), `GRG\_lp\_maj' (38.4\%), `GRG\_hp\_maj' (39.2\%), and 'GRG\_hp\_edge' (35.1\%), and are annotated in each subfigure alongside a scale bar for reference.}
\label{Fig:Polarization} 
\end{figure*}

In Fig.~\ref{Fig:Polarization}, we show the projected magnetic-field (B-field) lines (as inferred from the Stokes parameters; Section~\ref{Sec:Post-processing II: Generating polarization maps}) overlaid on the emission maps at their evolved times, with fractional polarization (FP) values proportional to the line lengths. An important observation from the maps is that, despite of our jet injection model (i.e., a purely toroidal B-field injected with the jet flow, initially perpendicular to the $x-$axis; Section~\ref{Sec:Revisiting Paper I}), the B-field lines eventually align with the flow direction (broadly along the $x-$axis), with added intricacies as the flow shows additional variations (e.g., bending, branching). This observed behavior aligns with the model proposed by \citet{Laing1981}, emphasizing the critical role of shock compression at the cocoon-ambient medium interface. Supplemented by the bulk flow, this process of compression increases shear and stress on the field lines, causing them to stretch and align tangentially to the lobe boundaries \citep[see, e.g.,][]{Roberts2008}.

In the low jet power cases in Fig.~\ref{Fig:Polarization}, a notable feature of the frontal lobes is the presence of lobe-sized vortices, where the bulk flow bends (counterclockwise in "GRG\_lp\_min" and clockwise in "GRG\_lp\_maj")$^{\ref{fn:Sim_Cases}}$. The projected B-field morphology closely follows the flow topology in these regions, reinforcing the conclusion that the jet material drags the magnetic field lines along with it as the flow progresses \citep[e.g.,][]{Baidoo2023}. The FP values remain high along the flows until vortex motion becomes dominant, inducing turbulence in both the flow and the magnetic field. This turbulence reduces the FP values in the projection, as reflected in the shorter line lengths (Fig.~\ref{Fig:Polarization}).

\begin{figure*}
\centering
\includegraphics[width=1.77\columnwidth]{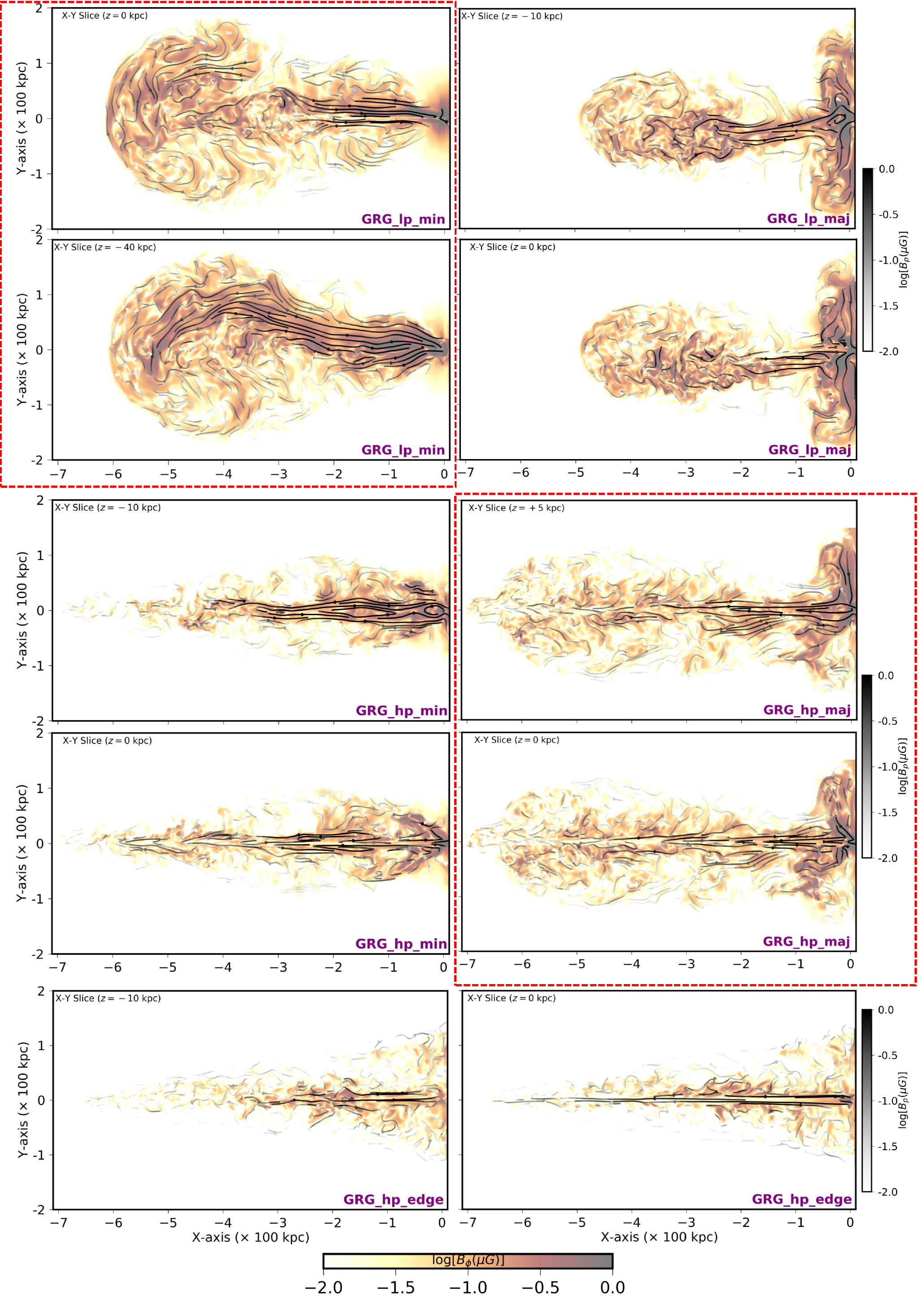}
\caption{The $x-y$ slices of the magnetic field configuration, taken at different $z-$plane positions. In each slice, the contours represent the poloidal magnetic field component ($B_p$), while the toroidal component ($B_{\phi}$) is shown using a colormap (in $\mu$G). Together, these components provide insight into (a) the dynamic characteristic of magnetic field structures within the jet–cocoon system and (b) their individual contribution to the features seen in polarization maps (Fig.~\ref{Fig:Polarization}). Additionally, the variation across $z$ demonstrates the non-axisymmetric bulk motion of the jet beam in three dimensions, emphasizing the limitations of employing a 2D model. To aid comparison of the field configurations within each group, red dashed lines have been added as visual guides to the reader.}
\label{Fig:Dynamic-Bmap} 
\end{figure*}

To correlate this observation further with the dynamical B-field distribution, we present 2D slices of magnetic fields from the 3D simulated data, in Fig.~\ref{Fig:Dynamic-Bmap}. In the figure, the contours represent the poloidal components of the B-field ($B_p$), while the colormap illustrates the toroidal component ($B_{\phi}$). The slices are chosen from different planes along the $z-$axis to showcase individual layer's contribution to the sky projected polarization map (note: the polarization maps in Fig.~\ref{Fig:Polarization} considers contribution of all cocoon layers, and so, the sliced dynamical maps can be treated only as a simplified interpretation), as indicated in the figure. In the low jet power cases, we observe that the dynamical B-field lines predominantly align along the flow direction. However, as smaller-scale vortices and turbulence layers dominate, both the strength and alignment of the field lines are significantly impacted (transforming into more chaotic arrangements).

A closer examination of the polarization maps also reveals the intricate filamentary structures within the lobes, as discussed in detail in Section~\ref{Sec:Correlating emission features with dynamical properties: Hotspot formation and filamentary extensions}. The projected B-field lines are observed to closely follow these filamentary trails, reaffirming that such internal filaments represent magnetic flows, resembling intertwined strands \citep[e.g.,][]{Sebokolodi2020,Andati2024}.

In the high jet power cases, a consistent pattern is observed where the B-field lines align strongly with the jet spine, extending nearly up to the jet head. However, as the distance increases, the stretching of the field lines gradually weakens the magnetic field. By comparing Fig.~\ref{Fig:Polarization} and \ref{Fig:Dynamic-Bmap}, it becomes clear that the B-field strength remains significant as long as the jet maintains its collimation. As the distance increases from jet injection, the backflow material becomes more dominant, carrying away part of the magnetic field. The generated backflow material also shows alignment of the field lines along the flow, but their turbulent nature becomes increasingly significant \citep[similar to,][]{Mukherjee2020}, particularly beyond the jet sheath structure, as prominently seen in the "GRG\_hp\_edge" case in both Fig.~\ref{Fig:Polarization} and \ref{Fig:Dynamic-Bmap}.

For the winged sources (the "maj" simulations), the magnetic field lines in the wings tend to align with the flow (fluid motion around the $y-$axis). The flow within these secondary lobes originate from the channeled backflowing plasma from the active lobe into the wing cavity, further reinforced by the backflowing matter from the opposing jet arm (discussed also in Section~\ref{Sec:Jet propagation through regions of maximal obstructions}). The B-field lines within the wings, observed to approximately aligning towards the y-axis in the projected polarization map, indicate the deflection and channelling of backflowing plasma into the wing. This configuration is particularly evident in the sliced dynamic maps in Fig.~\ref{Fig:Dynamic-Bmap}. At the head location of the wings, the compression of the B-field lines creates a bow-like arrangements. 

\begin{figure}
\centering
\includegraphics[width=\columnwidth]{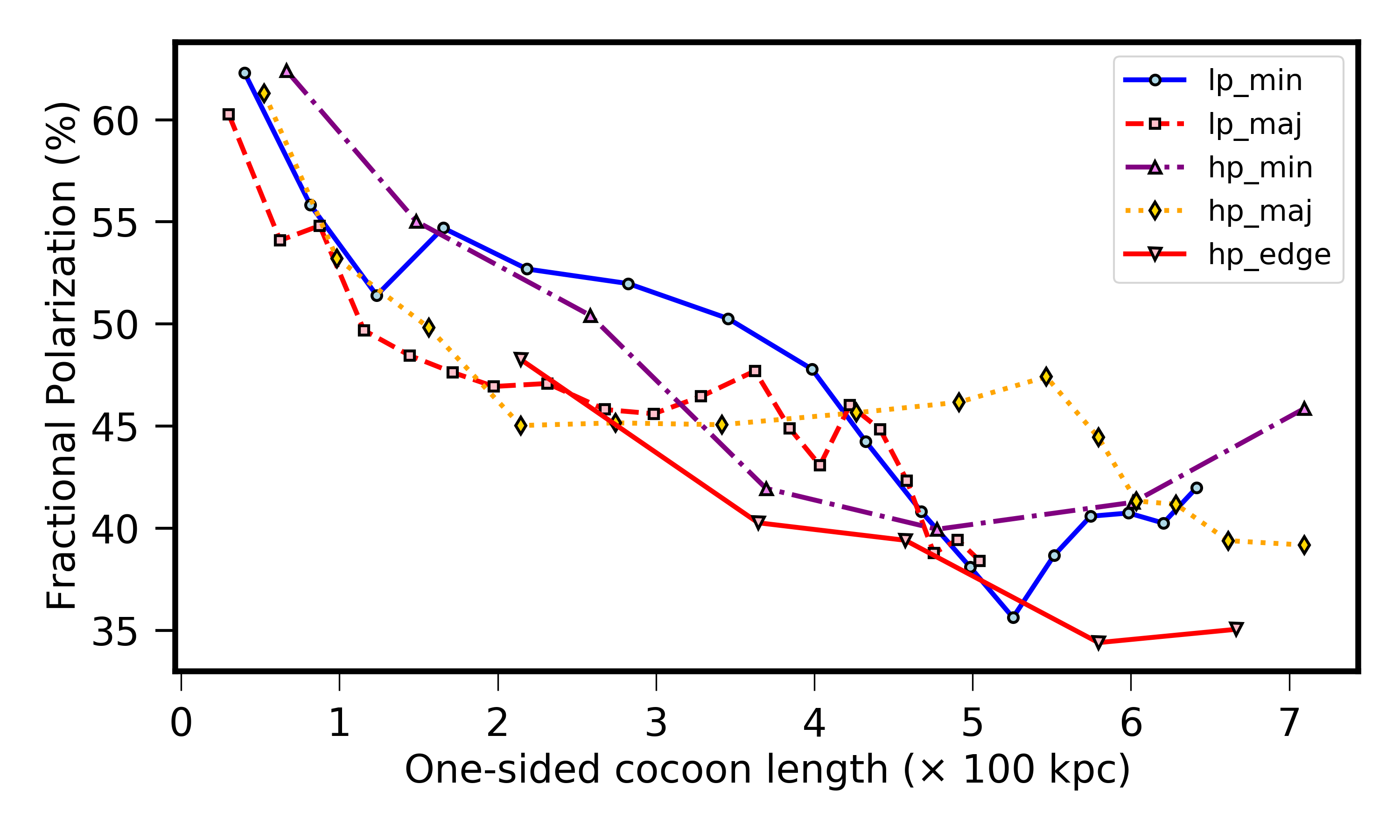}
\caption{Evolution of the median fractional polarization (FP) values in the simulated emission maps for the five situations, plotted against their growing length (mimicking also their aging), illustrating a gradual decrease in FP values followed by a near-settling phase. The FP values are calculated at 1 GHz at a spatial resolution of 1 kpc, corresponding to an angular resolution of 0.9 arcsecond for a redshift of 0.05.}
\label{Fig:Pol_vs_Len} 
\end{figure}

Building on the above discussion, which highlights the development of turbulence in the cocoon and its significant impact (diminishing) on FP estimates, we further analyzed the variation of FP values with the increasing length of the GRGs. This is presented in Fig.~\ref{Fig:Pol_vs_Len}, which shows a general decrease in the median FP values with growing length, suggesting that GRGs become increasingly turbulent in their cocoons as they evolve.
Initially, the FP values are relatively high, although still below the theoretical upper limit of $\sim$71\%, indicating that the jets carry an ordered magnetic field due to their high thrust (see also Fig.~\ref{Fig:Dynamic-Bmap}). However, as the GRG grows, the FP values gradually decline, showing a potential saturation at larger lengths, stabilizing around $\sim$40\%. See, \citet{Willis1978,Cotton2020} for relevance of these FP values in the context of GRGs. 

Whether this saturation indicates a settling phase in the evolution of GRGs or represents a brief transitional phase requires further investigation. Future studies, incorporating higher-resolution simulations and exploring the correlation length of the turbulent magnetic field distribution, will be essential to address this question. In that case, the effect of resolution on FP estimations can also be explored by convolving high-resolution maps to the beam sizes of different telescopes.

\section{Summary}\label{Sec:Summary}
The numerical modeling of giant radio galaxies and their long-term evolution has recently advanced, with the first 3D relativistic magnetohydrodynamic simulations presented in \citetalias{Giri2025} \citep{Giri2025}. \citetalias{Giri2025} focused on the intrinsic dynamical properties of GRGs, examining how their thermodynamic and magnetic characteristics evolve over age through a series of five simulations with varying jet powers and differing levels of resistance from the ambient medium to jet propagation. However, it is crucial to validate these results against observable quantities, particularly those measurable by contemporary radio telescopes. Such observations are key to addressing enduring questions in GRG physics, including the persistence of their emission despite the diminishing influence of several physical processes over time. Using a standard emission treatment to transform the 3D data of dynamical entities into sky-projected continuum and polarization maps at 1 GHz, we uncover the role of a few of these mechanisms intricately linked to the evolution of GRGs, as detailed below:

\begin{enumerate}
    \item In studying the matter transport processes within the GRG cocoon, we found that decollimation and bending are characteristic features of low-power jets, which lead to lobe formation. In contrast, high-power jets maintain their collimation, resulting in arrowhead- or nose-cone-shaped structures. This morphological distinction is additionally influenced by tri-axiality of the surrounding environment, giving rise to features such as winged sources and asymmetric cocoon extents. These findings thus suggest that GRG morphology can serve as a valuable proxy for disentangling the contributions of jet power and environmental factors.
    \item A prominent feature observable from the emission maps is the formation of large-scale vortex motion in low-powered GRGs and the branching or bifurcation of jet flows in high-powered GRGs. Both phenomena lead to decollimation and the generation of filamentary structures with varying geometries, such as tendril- or finger-like extensions and caldera-like formations, characterized by emission valleys surrounded by enhanced emission ridges, beside the formation of other smaller-scaled non-linear topologies.
    \item In the high-powered jet cases, prominent enhanced emission region often appears where the jet begins to lose its collimation, located approximately midway along the total cocoon extent. Despite identifying (weak) warm spots at the ends of two out of three high-powered simulations, the presence of such mid-cocoon enhanced emission zones could lead to misclassifications, suggesting that the GRG is restarting in nature (similar to concerns highlighted in recent numerical studies of smaller radio galaxies). In such scenarios, we propose studying the lateral intensity variations in GRGs (perpendicular to the bulk flow). An inactive or low-activity cocoon would exhibit a gradual decline in lateral emission, whereas active, high-powered GRGs would display abrupt variations in lateral intensity measures.
    \item A quantitative analysis of the five synthetic continuum maps reveals a set of key findings, as follows. First, applying a simple jet power–radio power conversion relationship to a statistical sample of GRGs can be misleading, as it varies with the parameters shaping individual GRG formation. Second, different estimates of magnetic field values show significant deviations from equipartition assumptions. Inverse Compton cooling by CMB photons appears to be a dominant cooling mechanism for GRGs, even at low redshifts. Lastly, matter entrainment has a diminishing role in sustaining GRG emission; low-powered jets entrain more environmental matter due to lobe formation, while high-powered jets retain collimation and thrust, limiting entrainment in the jet spine but heavily impacting the cocoon’s persistence.
    \item Striking similarities between the filamentary structures within the cocoon (as detected in the emission maps) and the Alfvénic velocity map, supported by aligned projected B-field lines and higher fractional polarization values along such threads in the polarization maps, confirm the magnetized nature of these internal filaments. These prominent filamentary structures within the cocoon, often identified in observed maps through simple edge-detection techniques, persisting for several Myr (dynamical age).
    \item Our findings suggest that the reduction in fractional polarization values is linked to the development of turbulence within the cocoon of GRGs. 
    As the jetted sources age and grow in length, the fractional polarization exhibits a diminishing trend, indicating a transition of the magnetic fields into more chaotic configurations. 
    Such a phenomenon can further be modeled   
    with varying resolutions (or beam sizes) and frequencies to determine the correlation length of the underlying turbulence. 
    Regardless of our jet injection models, the projected B-field lines align nearly parallel to the bulk flow of the jet-cocoon system.
\end{enumerate}

While this study serves as a foundational step to correlate the simulated dynamic and synthetic emission characteristics of GRGs, it provides a basis upon which future advancements and refinements can be achieved. First, a broader parameter set needs to be explored numerically to model the observed characteristics of GRGs in a statistical sample, including their prevalence as FR II GRGs with prominent hotspots at the jet head. This would involve exploring scenarios with a lower jet-to-environment density contrast, such as incorporating heavier jets, simulating jet propagation in a denser ambient medium, or injecting jets with higher power. Furthermore, examining diminishing effects, including matter entrainment in high-resolution simulation grids, its dependence on magnetic field strength, and the incorporation of radiative cooling into simulations, is crucial for understanding the persistence of emission during the longer-term evolution of GRGs. Several of these aspects were highlighted in \citetalias{Giri2025} and are planned for further investigation in our future work. Additionally, emphasis can be placed on translating these dynamical simulations into pragmatic radio maps, using more accurate emission conversion methods, such as the Lagrangian-Eulerian approach \citep[e.g.,][]{Vaidya2018}, which accounts for particle re-acceleration processes and the impact of turbulence on polarization behavior.

\begin{acknowledgements}
GG and KT acknowledge support from the South African Department of Science and Innovation's National Research Foundation under the ISARP RADIOMAP Joint Research Scheme (DSI-NRF Grant Number 150551).
CF acknowledges support by the German Research Foundation DFG via the research unit FORS\,5195. JB acknowledges the support from the Department of Physics and Electronics, Christ University, Bangalore. JD and GG acknowledges partial research-travel support by the National Research Foundation of South Africa (Ref Number CSUR240426216203). GG acknowledges the travel support from Gianluigi Bodo and Paola Rossi (INAF Torino) through Ob. Fu. 1.05.01.89.01 and Ricerca di base gruppo extragalattico, where discussions played a pivotal role in shaping this project. RPD acknowledges funding by the South African Research Chairs Initiative of the Department of Science, Technology, and Innovation and National Research Foundation (Grant ID: 77948). We acknowledge the use of the ilifu cloud computing facility – \url{https://www.ilifu.ac.za/}, a partnership between the University of Cape Town, the University of the Western Cape, Stellenbosch University, Sol Plaatje University, the Cape Peninsula University of Technology and the South African Radio Astronomy Observatory. The ilifu facility is supported by contributions from the InterUniversity Institute for Data Intensive Astronomy (IDIA – a partnership between the University of Cape Town, the University of Pretoria and the University of the Western Cape), the Computational Biology division at UCT and the Data Intensive Research Initiative of South Africa (DIRISA). The authors acknowledge the Centre for High Performance Computing (CHPC), South Africa, for providing computational resources to this research project (\url{https://www.chpc.ac.za/}).
\end{acknowledgements}

%
%


\bibliographystyle{aa} 
\bibliography{sample1,biblio} 

\begin{thebibliography}{126}
\expandafter\ifx\csname natexlab\endcsname\relax\def\natexlab#1{#1}\fi

\bibitem[{{Abolmasov} \& {Bromberg}(2023)}]{Abolmasov2023}
{Abolmasov}, P. \& {Bromberg}, O. 2023, \mnras, 520, 3009

\bibitem[{{Andati} {et~al.}(2024){Andati}, {Baidoo}, {Ramaila}, {Smirnov}, {Makhathini}, \& {Perley}}]{Andati2024}
{Andati}, L. A.~L., {Baidoo}, L.~M., {Ramaila}, A. J.~T., {et~al.} 2024, \mnras, 529, 1626

\bibitem[{{Andernach} {et~al.}(2021){Andernach}, {Jim{\'e}nez-Andrade}, \& {Willis}}]{Andernach2021}
{Andernach}, H., {Jim{\'e}nez-Andrade}, E.~F., \& {Willis}, A.~G. 2021, Galaxies, 9, 99

\bibitem[{{Bagchi} {et~al.}(2024){Bagchi}, {Ray}, {Dhiwar}, {Dabhade}, {Barth}, {Ho}, {Mirakhor}, {Walker}, {Nesvadba}, {Combes}, {Fabian}, \& {Jacob}}]{Bagchi2024}
{Bagchi}, J., {Ray}, S., {Dhiwar}, S., {et~al.} 2024, arXiv e-prints, arXiv:2405.01910

\bibitem[{{Bagchi} {et~al.}(2014){Bagchi}, {Vivek}, {Vikram}, {Hota}, {Biju}, {Sirothia}, {Srianand}, {Gopal-Krishna}, \& {Jacob}}]{2014ApJ...788..174B}
{Bagchi}, J., {Vivek}, M., {Vikram}, V., {et~al.} 2014, \apj, 788, 174

\bibitem[{{Baghel} {et~al.}(2023){Baghel}, {Kharb}, {Silpa}, {Ho}, \& {Harrison}}]{Baghel2023}
{Baghel}, J., {Kharb}, P., {Silpa}, {Ho}, L.~C., \& {Harrison}, C.~M. 2023, \mnras, 519, 2773

\bibitem[{{Baidoo} {et~al.}(2023){Baidoo}, {Perley}, {Eilek}, {Smirnov}, {Vacca}, \& {En{\ss}lin}}]{Baidoo2023}
{Baidoo}, L., {Perley}, R.~A., {Eilek}, J., {et~al.} 2023, \apj, 955, 16

\bibitem[{{Baldi}(2023)}]{Baldi2023}
{Baldi}, R.~D. 2023, \aapr, 31, 3

\bibitem[{{Black} {et~al.}(1992){Black}, {Baum}, {Leahy}, {Perley}, {Riley}, \& {Scheuer}}]{Black1992}
{Black}, A.~R.~S., {Baum}, S.~A., {Leahy}, J.~P., {et~al.} 1992, \mnras, 256, 186

\bibitem[{{Blandford} {et~al.}(2019){Blandford}, {Meier}, \& {Readhead}}]{Blandford2019}
{Blandford}, R., {Meier}, D., \& {Readhead}, A. 2019, \araa, 57, 467

\bibitem[{{Bodo} {et~al.}(2016){Bodo}, {Mamatsashvili}, {Rossi}, \& {Mignone}}]{Bodo2016}
{Bodo}, G., {Mamatsashvili}, G., {Rossi}, P., \& {Mignone}, A. 2016, \mnras, 462, 3031

\bibitem[{{Borse} {et~al.}(2021){Borse}, {Acharya}, {Vaidya}, {Mukherjee}, {Bodo}, {Rossi}, \& {Mignone}}]{Borse2021}
{Borse}, N., {Acharya}, S., {Vaidya}, B., {et~al.} 2021, \aap, 649, A150

\bibitem[{{Brienza} {et~al.}(2021){Brienza}, {Shimwell}, {de Gasperin}, {Bikmaev}, {Bonafede}, {Botteon}, {Br{\"u}ggen}, {Brunetti}, {Burenin}, {Capetti}, {Churazov}, {Hardcastle}, {Khabibullin}, {Lyskova}, {R{\"o}ttgering}, {Sunyaev}, {van Weeren}, {Gastaldello}, {Mandal}, {Purser}, {Simionescu}, \& {Tasse}}]{Brienza2021}
{Brienza}, M., {Shimwell}, T.~W., {de Gasperin}, F., {et~al.} 2021, Nature Astronomy, 5, 1261

\bibitem[{{Bromberg} {et~al.}(2011){Bromberg}, {Nakar}, {Piran}, \& {Sari}}]{Bromberg2011}
{Bromberg}, O., {Nakar}, E., {Piran}, T., \& {Sari}, R. 2011, \apj, 740, 100

\bibitem[{{Br{\"u}ggen} {et~al.}(2021){Br{\"u}ggen}, {Reiprich}, {Bulbul}, {Koribalski}, {Andernach}, {Rudnick}, {Hoang}, {Wilber}, {Duchesne}, {Veronica}, {Pacaud}, {Hopkins}, {Norris}, {Johnston-Hollitt}, {Brown}, {Bonafede}, {Brunetti}, {Collier}, {Sanders}, {Vardoulaki}, {Venturi}, {Kapinska}, \& {Marvil}}]{Bruggen2021}
{Br{\"u}ggen}, M., {Reiprich}, T.~H., {Bulbul}, E., {et~al.} 2021, \aap, 647, A3

\bibitem[{{Bruni} {et~al.}(2021){Bruni}, {Brienza}, {Panessa}, {Bassani}, {Dallacasa}, {Venturi}, {Baldi}, {Botteon}, {Drabent}, {Malizia}, {Massaro}, {R{\"o}ttgering}, {Ubertini}, {Ursini}, \& {van Weeren}}]{Bruni2021}
{Bruni}, G., {Brienza}, M., {Panessa}, F., {et~al.} 2021, \mnras, 503, 4681

\bibitem[{{Bruno} {et~al.}(2024){Bruno}, {Brienza}, {Zanichelli}, {Gitti}, {Ubertosi}, {Rajpurohit}, {Venturi}, \& {Dallacasa}}]{Bruno2024}
{Bruno}, L., {Brienza}, M., {Zanichelli}, A., {et~al.} 2024, \aap, 690, A160

\bibitem[{{Cantwell} {et~al.}(2020){Cantwell}, {Bray}, {Croston}, {Scaife}, {Mulcahy}, {Best}, {Br{\"u}ggen}, {Brunetti}, {Callingham}, {Clarke}, {Hardcastle}, {Harwood}, {Heald}, {Heesen}, {Iacobelli}, {Jamrozy}, {Morganti}, {Orr{\'u}}, {O'Sullivan}, {Riseley}, {R{\"o}ttgering}, {Shulevski}, {Sridhar}, {Tasse}, \& {Van Eck}}]{Cantwell2020}
{Cantwell}, T.~M., {Bray}, J.~D., {Croston}, J.~H., {et~al.} 2020, \mnras, 495, 143

\bibitem[{{Capetti} {et~al.}(2002){Capetti}, {Zamfir}, {Rossi}, {Bodo}, {Zanni}, \& {Massaglia}}]{Capetti2002}
{Capetti}, A., {Zamfir}, S., {Rossi}, P., {et~al.} 2002, \aap, 394, 39

\bibitem[{{Charlton} {et~al.}(2024){Charlton}, {Delhaize}, {Thorat}, {Heywood}, {Jarvis}, {Hardcastle}, {An}, {Delvecchio}, {Hale}, {Whittam}, {Br{\"u}ggen}, {Marchetti}, {Morabito}, {Randriamanakoto}, {White}, \& {Taylor}}]{Charlton2024}
{Charlton}, K.~K.~L., {Delhaize}, J., {Thorat}, K., {et~al.} 2024, arXiv e-prints, arXiv:2411.06813

\bibitem[{{Chen} {et~al.}(2018){Chen}, {Strom}, \& {Peng}}]{Chen2018}
{Chen}, R.-R., {Strom}, R., \& {Peng}, B. 2018, \apj, 858, 83

\bibitem[{{Cheung}(2007)}]{Cheung2007}
{Cheung}, C.~C. 2007, \aj, 133, 2097

\bibitem[{{Cielo} {et~al.}(2017){Cielo}, {Antonuccio-Delogu}, {Silk}, \& {Romeo}}]{Cielo2017}
{Cielo}, S., {Antonuccio-Delogu}, V., {Silk}, J., \& {Romeo}, A.~D. 2017, \mnras, 467, 4526

\bibitem[{{Clarke} {et~al.}(2017){Clarke}, {Heald}, {Jarrett}, {Bray}, {Hardcastle}, {Cantwell}, {Scaife}, {Brienza}, {Bonafede}, {Breton}, {Broderick}, {Carbone}, {Croston}, {Farnes}, {Harwood}, {Heesen}, {Horneffer}, {van der Horst}, {Iacobelli}, {Jurusik}, {Kokotanekov}, {McKean}, {Morabito}, {Mulcahy}, {Nikiel-Wroczy{\~n}ski}, {Orr{\'u}}, {Paladino}, {Pandey-Pommier}, {Pietka}, {Pizzo}, {Pratley}, {Riseley}, {Rottgering}, {Rowlinson}, {Sabater}, {Sendlinger}, {Shulevski}, {Sridhar}, {Stewart}, {Tasse}, {van Velzen}, {van Weeren}, \& {Wise}}]{Clarke2017}
{Clarke}, A.~O., {Heald}, G., {Jarrett}, T., {et~al.} 2017, \aap, 601, A25

\bibitem[{{Condon} {et~al.}(2021){Condon}, {Cotton}, {White}, {Legodi}, {Goedhart}, {McAlpine}, {Ratcliffe}, \& {Camilo}}]{Condon2021}
{Condon}, J.~J., {Cotton}, W.~D., {White}, S.~V., {et~al.} 2021, \apj, 917, 18

\bibitem[{{Cotton} {et~al.}(2025){Cotton}, {Giri}, {Agnihotri}, {Saikia}, {Thorat}, \& {Camilo}}]{Cotton2025}
{Cotton}, W.~D., {Giri}, G., {Agnihotri}, P.~J., {et~al.} 2025, \apj, 989, 223

\bibitem[{{Cotton} {et~al.}(2020){Cotton}, {Thorat}, {Condon}, {Frank}, {J{\'o}zsa}, {White}, {Deane}, {Oozeer}, {Atemkeng}, {Bester}, {Fanaroff}, {Kupa}, {Smirnov}, {Mauch}, {Krishnan}, \& {Camilo}}]{Cotton2020}
{Cotton}, W.~D., {Thorat}, K., {Condon}, J.~J., {et~al.} 2020, \mnras, 495, 1271

\bibitem[{{Dabhade} {et~al.}(2020){Dabhade}, {Mahato}, {Bagchi}, {Saikia}, {Combes}, {Sankhyayan}, {R{\"o}ttgering}, {Ho}, {Gaikwad}, {Raychaudhury}, {Vaidya}, \& {Guiderdoni}}]{Dabhade2020_SaganI}
{Dabhade}, P., {Mahato}, M., {Bagchi}, J., {et~al.} 2020, \aap, 642, A153

\bibitem[{{Dabhade} {et~al.}(2023){Dabhade}, {Saikia}, \& {Mahato}}]{Dabhade2023}
{Dabhade}, P., {Saikia}, D.~J., \& {Mahato}, M. 2023, Journal of Astrophysics and Astronomy, 44, 13

\bibitem[{{Dav{\'e}} {et~al.}(2001){Dav{\'e}}, {Cen}, {Ostriker}, {Bryan}, {Hernquist}, {Katz}, {Weinberg}, {Norman}, \& {O'Shea}}]{Dave2001}
{Dav{\'e}}, R., {Cen}, R., {Ostriker}, J.~P., {et~al.} 2001, \apj, 552, 473

\bibitem[{{Del Zanna} {et~al.}(2006){Del Zanna}, {Volpi}, {Amato}, \& {Bucciantini}}]{DelZanna2006}
{Del Zanna}, L., {Volpi}, D., {Amato}, E., \& {Bucciantini}, N. 2006, \aap, 453, 621

\bibitem[{{Delhaize} {et~al.}(2021){Delhaize}, {Heywood}, {Prescott}, {Jarvis}, {Delvecchio}, {Whittam}, {White}, {Hardcastle}, {Hale}, {Afonso}, {Ao}, {Brienza}, {Br{\"u}ggen}, {Collier}, {Daddi}, {Glowacki}, {Maddox}, {Morabito}, {Prandoni}, {Randriamanakoto}, {Sekhar}, {An}, {Adams}, {Blyth}, {Bowler}, {Leeuw}, {Marchetti}, {Randriamampandry}, {Thorat}, {Seymour}, {Smirnov}, {Taylor}, {Tasse}, \& {Vaccari}}]{Delhaize2021}
{Delhaize}, J., {Heywood}, I., {Prescott}, M., {et~al.} 2021, \mnras, 501, 3833

\bibitem[{{Duan} {et~al.}(2024){Duan}, {Wu}, {Zhang}, \& {Li}}]{Duan2024}
{Duan}, X., {Wu}, L., {Zhang}, R., \& {Li}, J. 2024, arXiv e-prints, arXiv:2410.04467

\bibitem[{{Dubey} {et~al.}(2023){Dubey}, {Fendt}, \& {Vaidya}}]{Dubey2023}
{Dubey}, R.~P., {Fendt}, C., \& {Vaidya}, B. 2023, \apj, 952, 1

\bibitem[{{Dubey} {et~al.}(2024){Dubey}, {Fendt}, \& {Vaidya}}]{Dubey2024}
{Dubey}, R.~P., {Fendt}, C., \& {Vaidya}, B. 2024, \apj, 976, 144

\bibitem[{{Fanaroff} {et~al.}(2021){Fanaroff}, {Lal}, {Venturi}, {Smirnov}, {Bondi}, {Thorat}, {Bester}, {J{\'o}zsa}, {Kleiner}, {Loi}, {Makhathini}, \& {White}}]{Fanaroff2021}
{Fanaroff}, B., {Lal}, D.~V., {Venturi}, T., {et~al.} 2021, \mnras, 505, 6003

\bibitem[{{Fanaroff} \& {Riley}(1974)}]{Fanaroff1974}
{Fanaroff}, B.~L. \& {Riley}, J.~M. 1974, \mnras, 167, 31P

\bibitem[{{Giri} {et~al.}(2025){Giri}, {Bagchi}, {Thorat}, {Deane}, {Delhaize}, \& {Saikia}}]{Giri2025}
{Giri}, G., {Bagchi}, J., {Thorat}, K., {et~al.} 2025, \aap, 693, A77

\bibitem[{{Giri} {et~al.}(2022{\natexlab{a}}){Giri}, {Dubey}, {Rubinur}, {Vaidya}, \& {Kharb}}]{Giri2022_S}
{Giri}, G., {Dubey}, R.~P., {Rubinur}, K., {Vaidya}, B., \& {Kharb}, P. 2022{\natexlab{a}}, \mnras, 514, 5625

\bibitem[{{Giri} {et~al.}(2024){Giri}, {Fendt}, {Thorat}, {Bodo}, \& {Rossi}}]{Giri2024_Review}
{Giri}, G., {Fendt}, C., {Thorat}, K., {Bodo}, G., \& {Rossi}, P. 2024, Frontiers in Astronomy and Space Sciences, 11, 1371101

\bibitem[{{Giri} {et~al.}(2023){Giri}, {Vaidya}, \& {Fendt}}]{Giri2023}
{Giri}, G., {Vaidya}, B., \& {Fendt}, C. 2023, \apjs, 268, 49

\bibitem[{{Giri} {et~al.}(2022{\natexlab{b}}){Giri}, {Vaidya}, {Rossi}, {Bodo}, {Mukherjee}, \& {Mignone}}]{Giri2022_X}
{Giri}, G., {Vaidya}, B., {Rossi}, P., {et~al.} 2022{\natexlab{b}}, \aap, 662, A5

\bibitem[{{Gopal Krishna} \& {Dabhade}(2022)}]{Gopal-Krishna2022}
{Gopal Krishna} \& {Dabhade}, P. 2022, \aap, 663, L8

\bibitem[{{Gourgouliatos} \& {Komissarov}(2018)}]{Gourgouliatos2018}
{Gourgouliatos}, K.~N. \& {Komissarov}, S.~S. 2018, \mnras, 475, L125

\bibitem[{{Hardcastle}(2018)}]{Hardcastle2018}
{Hardcastle}, M.~J. 2018, \mnras, 475, 2768

\bibitem[{{Hardcastle} {et~al.}(2002){Hardcastle}, {Birkinshaw}, {Cameron}, {Harris}, {Looney}, \& {Worrall}}]{Hardcastle2002}
{Hardcastle}, M.~J., {Birkinshaw}, M., {Cameron}, R.~A., {et~al.} 2002, \apj, 581, 948

\bibitem[{{Hardcastle} \& {Croston}(2020)}]{Hardcastle2020}
{Hardcastle}, M.~J. \& {Croston}, J.~H. 2020, \nar, 88, 101539

\bibitem[{{Hardcastle} {et~al.}(2019){Hardcastle}, {Croston}, {Shimwell}, {Tasse}, {G{\"u}rkan}, {Morganti}, {Murgia}, {R{\"o}ttgering}, {van Weeren}, \& {Williams}}]{Hardcastle2019}
{Hardcastle}, M.~J., {Croston}, J.~H., {Shimwell}, T.~W., {et~al.} 2019, \mnras, 488, 3416

\bibitem[{{Hazard} {et~al.}(1963){Hazard}, {Mackey}, \& {Shimmins}}]{Hazard1963}
{Hazard}, C., {Mackey}, M.~B., \& {Shimmins}, A.~J. 1963, \nat, 197, 1037

\bibitem[{{Horton} {et~al.}(2023){Horton}, {Krause}, \& {Hardcastle}}]{Horton2023}
{Horton}, M.~A., {Krause}, M. G.~H., \& {Hardcastle}, M.~J. 2023, \mnras, 521, 2593

\bibitem[{{Ishwara-Chandra} \& {Saikia}(1999)}]{Ishwara-Chandra1999}
{Ishwara-Chandra}, C.~H. \& {Saikia}, D.~J. 1999, \mnras, 309, 100

\bibitem[{{Jamrozy} {et~al.}(2008){Jamrozy}, {Konar}, {Machalski}, \& {Saikia}}]{Jamrozy2008}
{Jamrozy}, M., {Konar}, C., {Machalski}, J., \& {Saikia}, D.~J. 2008, \mnras, 385, 1286

\bibitem[{{Jennison} \& {Das Gupta}(1953)}]{Jennison1953}
{Jennison}, R.~C. \& {Das Gupta}, M.~K. 1953, \nat, 172, 996

\bibitem[{{Knowles} {et~al.}(2022){Knowles}, {Cotton}, {Rudnick}, {Camilo}, {Goedhart}, {Deane}, {Ramatsoku}, {Bietenholz}, {Br{\"u}ggen}, {Button}, {Chen}, {Chibueze}, {Clarke}, {de Gasperin}, {Ianjamasimanana}, {J{\'o}zsa}, {Hilton}, {Kesebonye}, {Kolokythas}, {Kraan-Korteweg}, {Lawrie}, {Lochner}, {Loubser}, {Marchegiani}, {Mhlahlo}, {Moodley}, {Murphy}, {Namumba}, {Oozeer}, {Parekh}, {Pillay}, {Passmoor}, {Ramaila}, {Ranchod}, {Retana-Montenegro}, {Sebokolodi}, {Sikhosana}, {Smirnov}, {Thorat}, {Venturi}, {Abbott}, {Adam}, {Adams}, {Aldera}, {Bauermeister}, {Bennett}, {Bode}, {Botha}, {Botha}, {Brederode}, {Buchner}, {Burger}, {Cheetham}, {de Villiers}, {Dikgale-Mahlakoana}, {du Toit}, {Esterhuyse}, {Fadana}, {Fanaroff}, {Fataar}, {Foley}, {Fourie}, {Frank}, {Gamatham}, {Gatsi}, {Geyer}, {Gouws}, {Gumede}, {Heywood}, {Hlakola}, {Hokwana}, {Hoosen}, {Horn}, {Horrell}, {Hugo}, {Isaacson}, {Jonas}, {Jordaan}, {Joubert}, {Julie}, {Kapp}, {Kasper}, {Kenyon}, {Kotz{\'e}}, {Kotze}, {Kriek}, {Kriel}, {Krishnan},
  {Kusel}, {Legodi}, {Lehmensiek}, {Liebenberg}, {Lord}, {Lunsky}, {Madisa}, {Magnus}, {Main}, {Makhaba}, {Makhathini}, {Malan}, {Manley}, {Marais}, {Maree}, {Martens}, {Mauch}, {McAlpine}, {Merry}, {Millenaar}, {Mokone}, {Monama}, {Mphego}, {New}, {Ngcebetsha}, {Ngoasheng}, {Ockards}, {Otto}, {Patel}, {Peens-Hough}, {Perkins}, {Ramanujam}, {Ramudzuli}, {Ratcliffe}, {Renil}, {Robyntjies}, {Rust}, {Salie}, {Sambu}, {Schollar}, {Schwardt}, {Schwartz}, {Serylak}, {Siebrits}, {Sirothia}, {Slabber}, {Sofeya}, {Taljaard}, {Tasse}, {Tiplady}, {Toruvanda}, {Twum}, {van Balla}, {van der Byl}, {van der Merwe}, {van Dyk}, {Van Tonder}, {Van Wyk}, {Venter}, {Venter}, {Welz}, {Williams}, \& {Xaia}}]{Knowles2022}
{Knowles}, K., {Cotton}, W.~D., {Rudnick}, L., {et~al.} 2022, \aap, 657, A56

\bibitem[{{Komberg} \& {Pashchenko}(2009)}]{Komberg2009}
{Komberg}, B.~V. \& {Pashchenko}, I.~N. 2009, Astronomy Reports, 53, 1086

\bibitem[{{Koribalski}(2025)}]{Koribalski2025}
{Koribalski}, B.~S. 2025, arXiv e-prints, arXiv:2504.07314

\bibitem[{{Koribalski} {et~al.}(2024){Koribalski}, {Duchesne}, {Lenc}, {Venturi}, {Botteon}, {Shabala}, {Vernstrom}, {Carretti}, {Norris}, {Anderson}, {Hopkins}, {Riseley}, {Gupta}, \& {Velovi{\'c}}}]{Koribalski2024}
{Koribalski}, B.~S., {Duchesne}, S.~W., {Lenc}, E., {et~al.} 2024, \mnras, 533, 608

\bibitem[{{Kundu} {et~al.}(2022){Kundu}, {Vaidya}, {Mignone}, \& {Hardcastle}}]{Kundu2022}
{Kundu}, S., {Vaidya}, B., {Mignone}, A., \& {Hardcastle}, M.~J. 2022, \aap, 667, A138

\bibitem[{{Laing}(1981)}]{Laing1981}
{Laing}, R.~A. 1981, \apj, 248, 87

\bibitem[{{Laing} \& {Bridle}(2012)}]{Laing2012}
{Laing}, R.~A. \& {Bridle}, A.~H. 2012, \mnras, 424, 1149

\bibitem[{{Laing} \& {Bridle}(2015)}]{Laing2015}
{Laing}, R.~A. \& {Bridle}, A.~H. 2015, in IAU Symposium, Vol. 313, Extragalactic Jets from Every Angle, ed. F.~{Massaro}, C.~C. {Cheung}, E.~{Lopez}, \& A.~{Siemiginowska}, 108--115

\bibitem[{{Leahy} {et~al.}(1997){Leahy}, {Black}, {Dennett-Thorpe}, {Hardcastle}, {Komissarov}, {Perley}, {Riley}, \& {Scheuer}}]{Leahy1997}
{Leahy}, J.~P., {Black}, A.~R.~S., {Dennett-Thorpe}, J., {et~al.} 1997, \mnras, 291, 20

\bibitem[{{Machalski}(2011)}]{Machalski2011}
{Machalski}, J. 2011, \mnras, 413, 2429

\bibitem[{{Machalski} {et~al.}(2009){Machalski}, {Jamrozy}, \& {Saikia}}]{Machalski2009}
{Machalski}, J., {Jamrozy}, M., \& {Saikia}, D.~J. 2009, \mnras, 395, 812

\bibitem[{{Machalski} {et~al.}(2008){Machalski}, {Kozie{\l}-Wierzbowska}, {Jamrozy}, \& {Saikia}}]{Machalski2008}
{Machalski}, J., {Kozie{\l}-Wierzbowska}, D., {Jamrozy}, M., \& {Saikia}, D.~J. 2008, \apj, 679, 149

\bibitem[{{Mack} {et~al.}(1997){Mack}, {Klein}, {O'Dea}, \& {Willis}}]{Mack1997}
{Mack}, K.~H., {Klein}, U., {O'Dea}, C.~P., \& {Willis}, A.~G. 1997, \aaps, 123, 423

\bibitem[{{Mahatma}(2023)}]{Mahatma2023}
{Mahatma}, V.~H. 2023, Galaxies, 11, 74

\bibitem[{{Mahatma} {et~al.}(2020){Mahatma}, {Hardcastle}, {Croston}, {Harwood}, {Ineson}, \& {Moldon}}]{Mahatma2020}
{Mahatma}, V.~H., {Hardcastle}, M.~J., {Croston}, J.~H., {et~al.} 2020, \mnras, 491, 5015

\bibitem[{{Malarecki} {et~al.}(2015){Malarecki}, {Jones}, {Saripalli}, {Staveley-Smith}, \& {Subrahmanyan}}]{Malarecki2015}
{Malarecki}, J.~M., {Jones}, D.~H., {Saripalli}, L., {Staveley-Smith}, L., \& {Subrahmanyan}, R. 2015, \mnras, 449, 955

\bibitem[{{Malarecki} {et~al.}(2013){Malarecki}, {Staveley-Smith}, {Saripalli}, {Subrahmanyan}, {Jones}, {Duffy}, \& {Rioja}}]{Malarecki2013}
{Malarecki}, J.~M., {Staveley-Smith}, L., {Saripalli}, L., {et~al.} 2013, \mnras, 432, 200

\bibitem[{{Marecki} {et~al.}(2021){Marecki}, {Jamrozy}, {Machalski}, \& {Pajdosz-{\'S}mierciak}}]{Marecki2021}
{Marecki}, A., {Jamrozy}, M., {Machalski}, J., \& {Pajdosz-{\'S}mierciak}, U. 2021, \mnras, 501, 853

\bibitem[{{Massaglia} {et~al.}(2016){Massaglia}, {Bodo}, {Rossi}, {Capetti}, \& {Mignone}}]{Massaglia2016}
{Massaglia}, S., {Bodo}, G., {Rossi}, P., {Capetti}, S., \& {Mignone}, A. 2016, \aap, 596, A12

\bibitem[{{Matthews} {et~al.}(2019){Matthews}, {Bell}, {Blundell}, \& {Araudo}}]{Matthews2019}
{Matthews}, J.~H., {Bell}, A.~R., {Blundell}, K.~M., \& {Araudo}, A.~T. 2019, \mnras, 482, 4303

\bibitem[{{McKinley} {et~al.}(2021){McKinley}, {Tingay}, {Gaspari}, {Kraft}, {Matherne}, {Offringa}, {McDonald}, {Calzadilla}, {Veilleux}, {Shabala}, {Gwyn}, {Bland-Hawthorn}, {Crnojevic}, {Gaensler}, \& {Johnston-Hollitt}}]{McKinley:2021}
{McKinley}, B., {Tingay}, S.~J., {Gaspari}, M., {et~al.} 2021, Nature Astronomy

\bibitem[{{Meenakshi} {et~al.}(2023){Meenakshi}, {Mukherjee}, {Bodo}, \& {Rossi}}]{Meenakshi2023}
{Meenakshi}, M., {Mukherjee}, D., {Bodo}, G., \& {Rossi}, P. 2023, \mnras, 526, 5418

\bibitem[{{Mignone} {et~al.}(2005){Mignone}, {Plewa}, \& {Bodo}}]{Mignone2005}
{Mignone}, A., {Plewa}, T., \& {Bodo}, G. 2005, \apjs, 160, 199

\bibitem[{{Mignone} {et~al.}(2010){Mignone}, {Rossi}, {Bodo}, {Ferrari}, \& {Massaglia}}]{Mignone2010}
{Mignone}, A., {Rossi}, P., {Bodo}, G., {Ferrari}, A., \& {Massaglia}, S. 2010, \mnras, 402, 7

\bibitem[{{Mingo} {et~al.}(2019){Mingo}, {Croston}, {Hardcastle}, {Best}, {Duncan}, {Morganti}, {Rottgering}, {Sabater}, {Shimwell}, {Williams}, {Brienza}, {Gurkan}, {Mahatma}, {Morabito}, {Prandoni}, {Bondi}, {Ineson}, \& {Mooney}}]{Mingo2019}
{Mingo}, B., {Croston}, J.~H., {Hardcastle}, M.~J., {et~al.} 2019, \mnras, 488, 2701

\bibitem[{{Monceau-Baroux} {et~al.}(2015){Monceau-Baroux}, {Porth}, {Meliani}, \& {Keppens}}]{Monceau-Baroux2015}
{Monceau-Baroux}, R., {Porth}, O., {Meliani}, Z., \& {Keppens}, R. 2015, \aap, 574, A143

\bibitem[{{Mukherjee} {et~al.}(2020){Mukherjee}, {Bodo}, {Mignone}, {Rossi}, \& {Vaidya}}]{Mukherjee2020}
{Mukherjee}, D., {Bodo}, G., {Mignone}, A., {Rossi}, P., \& {Vaidya}, B. 2020, \mnras, 499, 681

\bibitem[{{Mukherjee} {et~al.}(2021){Mukherjee}, {Bodo}, {Rossi}, {Mignone}, \& {Vaidya}}]{Mukherjee2021}
{Mukherjee}, D., {Bodo}, G., {Rossi}, P., {Mignone}, A., \& {Vaidya}, B. 2021, \mnras, 505, 2267

\bibitem[{{Nawaz} {et~al.}(2016){Nawaz}, {Bicknell}, {Wagner}, {Sutherland}, \& {McNamara}}]{Nawaz2016}
{Nawaz}, M.~A., {Bicknell}, G.~V., {Wagner}, A.~Y., {Sutherland}, R.~S., \& {McNamara}, B.~R. 2016, \mnras, 458, 802

\bibitem[{{O'Dea} \& {Baum}(1997)}]{Odea1997}
{O'Dea}, C.~P. \& {Baum}, S.~A. 1997, \aj, 113, 148

\bibitem[{{O'Dea} \& {Saikia}(2021)}]{Odea2021}
{O'Dea}, C.~P. \& {Saikia}, D.~J. 2021, \aapr, 29, 3

\bibitem[{{Oei} {et~al.}(2024{\natexlab{a}}){Oei}, {Hardcastle}, {Timmerman}, {Gast}, {Botteon}, {Rodriguez}, {Stern}, {Calistro Rivera}, {van Weeren}, {R{\"o}ttgering}, {Intema}, {de Gasperin}, \& {Djorgovski}}]{Oei2024_7mpc}
{Oei}, M. S.~S.~L., {Hardcastle}, M.~J., {Timmerman}, R., {et~al.} 2024{\natexlab{a}}, \nat, 633, 537

\bibitem[{{Oei} {et~al.}(2023){Oei}, {van Weeren}, {Gast}, {Botteon}, {Hardcastle}, {Dabhade}, {Shimwell}, {R{\"o}ttgering}, \& {Drabent}}]{Oei2023_length_dist}
{Oei}, M. S.~S.~L., {van Weeren}, R.~J., {Gast}, A. R.~D.~J.~G.~I.~B., {et~al.} 2023, \aap, 672, A163

\bibitem[{{Oei} {et~al.}(2022){Oei}, {van Weeren}, {Hardcastle}, {Botteon}, {Shimwell}, {Dabhade}, {Gast}, {R{\"o}ttgering}, {Br{\"u}ggen}, {Tasse}, {Williams}, \& {Shulevski}}]{Oei2022_5Mpc}
{Oei}, M. S.~S.~L., {van Weeren}, R.~J., {Hardcastle}, M.~J., {et~al.} 2022, \aap, 660, A2

\bibitem[{{Oei} {et~al.}(2024{\natexlab{b}}){Oei}, {van Weeren}, {Hardcastle}, {Gast}, {Leclercq}, {R{\"o}ttgering}, {Dabhade}, {Shimwell}, \& {Botteon}}]{Oei2024_filament}
{Oei}, M. S.~S.~L., {van Weeren}, R.~J., {Hardcastle}, M.~J., {et~al.} 2024{\natexlab{b}}, \aap, 686, A137

\bibitem[{{Ogrodnik} {et~al.}(2021){Ogrodnik}, {Hanasz}, \& {W{\'o}lta{\'n}ski}}]{Ogrodnik2021}
{Ogrodnik}, M.~A., {Hanasz}, M., \& {W{\'o}lta{\'n}ski}, D. 2021, \apjs, 253, 18

\bibitem[{{Patra} {et~al.}(2023){Patra}, {Joshi}, \& {Gopal-Krishna}}]{Patra2023}
{Patra}, D., {Joshi}, R., \& {Gopal-Krishna}. 2023, \mnras, 524, 3270

\bibitem[{{Pe'er}(2014)}]{Peer2014}
{Pe'er}, A. 2014, \ssr, 183, 371

\bibitem[{{Pyrzas} {et~al.}(2015){Pyrzas}, {Steenbrugge}, \& {Blundell}}]{Pyrzas2015}
{Pyrzas}, S., {Steenbrugge}, K.~C., \& {Blundell}, K.~M. 2015, \aap, 574, A30

\bibitem[{{Ramatsoku} {et~al.}(2020){Ramatsoku}, {Murgia}, {Vacca}, {Serra}, {Makhathini}, {Govoni}, {Smirnov}, {Andati}, {de Blok}, {J{\'o}zsa}, {Kamphuis}, {Kleiner}, {Maccagni}, {Moln{\'a}r}, {Ramaila}, {Thorat}, \& {White}}]{Ramatsoku2020}
{Ramatsoku}, M., {Murgia}, M., {Vacca}, V., {et~al.} 2020, \aap, 636, L1

\bibitem[{{Roberts} {et~al.}(2008){Roberts}, {Wardle}, {Lipnick}, {Selesnick}, \& {Slutsky}}]{Roberts2008}
{Roberts}, D.~H., {Wardle}, J. F.~C., {Lipnick}, S.~L., {Selesnick}, P.~L., \& {Slutsky}, S. 2008, \apj, 676, 584

\bibitem[{{Rossi} {et~al.}(2017){Rossi}, {Bodo}, {Capetti}, \& {Massaglia}}]{Rossi2017}
{Rossi}, P., {Bodo}, G., {Capetti}, A., \& {Massaglia}, S. 2017, \aap, 606, A57

\bibitem[{{Rossi} {et~al.}(2024){Rossi}, {Bodo}, {Massaglia}, \& {Capetti}}]{Rossi2024}
{Rossi}, P., {Bodo}, G., {Massaglia}, S., \& {Capetti}, A. 2024, \aap, 685, A4

\bibitem[{{Rossi} {et~al.}(2008){Rossi}, {Mignone}, {Bodo}, {Massaglia}, \& {Ferrari}}]{Rossi2008}
{Rossi}, P., {Mignone}, A., {Bodo}, G., {Massaglia}, S., \& {Ferrari}, A. 2008, \aap, 488, 795

\bibitem[{{Rudnick} {et~al.}(2022){Rudnick}, {Br{\"u}ggen}, {Brunetti}, {Cotton}, {Forman}, {Jones}, {Nolting}, {Schellenberger}, \& {van Weeren}}]{Rudnick2022}
{Rudnick}, L., {Br{\"u}ggen}, M., {Brunetti}, G., {et~al.} 2022, \apj, 935, 168

\bibitem[{{Rudnick} {et~al.}(2021){Rudnick}, {Cotton}, {Knowles}, \& {Kolokythas}}]{Rudnick2021}
{Rudnick}, L., {Cotton}, W., {Knowles}, K., \& {Kolokythas}, K. 2021, Galaxies, 9, 81

\bibitem[{{Rybicki} \& {Lightman}(1979)}]{Rybicki1979}
{Rybicki}, G.~B. \& {Lightman}, A.~P. 1979, {Radiative processes in astrophysics}

\bibitem[{{Safouris} {et~al.}(2009){Safouris}, {Subrahmanyan}, {Bicknell}, \& {Saripalli}}]{Safouris2009}
{Safouris}, V., {Subrahmanyan}, R., {Bicknell}, G.~V., \& {Saripalli}, L. 2009, \mnras, 393, 2

\bibitem[{{Saikia}(2022)}]{Saikia2022}
{Saikia}, D.~J. 2022, Journal of Astrophysics and Astronomy, 43, 97

\bibitem[{{Saripalli} \& {Subrahmanyan}(2009)}]{Saripalli2009}
{Saripalli}, L. \& {Subrahmanyan}, R. 2009, \apj, 695, 156

\bibitem[{{Schoenmakers} {et~al.}(2000){Schoenmakers}, {Mack}, {de Bruyn}, {R{\"o}ttgering}, {Klein}, \& {van der Laan}}]{Schoenmakers2000}
{Schoenmakers}, A.~P., {Mack}, K.~H., {de Bruyn}, A.~G., {et~al.} 2000, \aaps, 146, 293

\bibitem[{{Sebastian} {et~al.}(2018){Sebastian}, {Ishwara-Chandra}, {Joshi}, \& {Wadadekar}}]{Sebastian2018}
{Sebastian}, B., {Ishwara-Chandra}, C.~H., {Joshi}, R., \& {Wadadekar}, Y. 2018, \mnras, 473, 4926

\bibitem[{{Sebokolodi} {et~al.}(2020){Sebokolodi}, {Perley}, {Eilek}, {Carilli}, {Smirnov}, {Laing}, {Greisen}, \& {Wise}}]{Sebokolodi2020}
{Sebokolodi}, M. L.~L., {Perley}, R., {Eilek}, J., {et~al.} 2020, \apj, 903, 36

\bibitem[{{Simonte} {et~al.}(2024){Simonte}, {Andernach}, {Br{\"u}ggen}, {Miley}, \& {Barthel}}]{Simonte2024}
{Simonte}, M., {Andernach}, H., {Br{\"u}ggen}, M., {Miley}, G.~K., \& {Barthel}, P. 2024, \aap, 686, A21

\bibitem[{{Simonte} {et~al.}(2022){Simonte}, {Andernach}, {Br{\"u}ggen}, {Schwarz}, {Prandoni}, \& {Willis}}]{Simonte2022}
{Simonte}, M., {Andernach}, H., {Br{\"u}ggen}, M., {et~al.} 2022, \mnras, 515, 2032

\bibitem[{{Stone} \& {Hardee}(2000)}]{STone2000}
{Stone}, J.~M. \& {Hardee}, P.~E. 2000, \apj, 540, 192

\bibitem[{{Stuardi} {et~al.}(2020){Stuardi}, {O'Sullivan}, {Bonafede}, {Br{\"u}ggen}, {Dabhade}, {Horellou}, {Morganti}, {Carretti}, {Heald}, {Iacobelli}, \& {Vacca}}]{Stuardi2020}
{Stuardi}, C., {O'Sullivan}, S.~P., {Bonafede}, A., {et~al.} 2020, \aap, 638, A48

\bibitem[{{Subrahmanyan} {et~al.}(1996){Subrahmanyan}, {Saripalli}, \& {Hunstead}}]{Subrahmanyan1996}
{Subrahmanyan}, R., {Saripalli}, L., \& {Hunstead}, R.~W. 1996, \mnras, 279, 257

\bibitem[{{Subrahmanyan} {et~al.}(2008){Subrahmanyan}, {Saripalli}, {Safouris}, \& {Hunstead}}]{Subrahmanyan2008}
{Subrahmanyan}, R., {Saripalli}, L., {Safouris}, V., \& {Hunstead}, R.~W. 2008, \apj, 677, 63

\bibitem[{{Tamhane} {et~al.}(2015){Tamhane}, {Wadadekar}, {Basu}, {Singh}, {Ishwara-Chandra}, {Beelen}, \& {Sirothia}}]{Tamhane2015}
{Tamhane}, P., {Wadadekar}, Y., {Basu}, A., {et~al.} 2015, \mnras, 453, 2438

\bibitem[{{Taub}(1948)}]{Taub1948}
{Taub}, A.~H. 1948, Physical Review, 74, 328

\bibitem[{{Turner} {et~al.}(2018){Turner}, {Rogers}, {Shabala}, \& {Krause}}]{Turner2018}
{Turner}, R.~J., {Rogers}, J.~G., {Shabala}, S.~S., \& {Krause}, M. G.~H. 2018, \mnras, 473, 4179

\bibitem[{{Ubertosi} {et~al.}(2025){Ubertosi}, {Gong}, {Nulsen}, {Leahy}, {Gitti}, {McNamara}, {Gaspari}, {Singha}, {O'Dea}, \& {Baum}}]{Ubertosi:2025}
{Ubertosi}, F., {Gong}, Y., {Nulsen}, P., {et~al.} 2025, \aap, 693, A171

\bibitem[{{Upreti} {et~al.}(2024){Upreti}, {Vaidya}, \& {Shukla}}]{Upreti2024}
{Upreti}, N., {Vaidya}, B., \& {Shukla}, A. 2024, Journal of High Energy Astrophysics, 44, 146

\bibitem[{{Ursini} {et~al.}(2018){Ursini}, {Bassani}, {Panessa}, {Bird}, {Bruni}, {Fiocchi}, {Malizia}, {Saripalli}, \& {Ubertini}}]{Ursini2018}
{Ursini}, F., {Bassani}, L., {Panessa}, F., {et~al.} 2018, \mnras, 481, 4250

\bibitem[{{Vaidya} {et~al.}(2018){Vaidya}, {Mignone}, {Bodo}, {Rossi}, \& {Massaglia}}]{Vaidya2018}
{Vaidya}, B., {Mignone}, A., {Bodo}, G., {Rossi}, P., \& {Massaglia}, S. 2018, \apj, 865, 144

\bibitem[{{van der Laan} \& {Perola}(1969)}]{VanDerLaan1969}
{van der Laan}, H. \& {Perola}, G.~C. 1969, \aap, 3, 468

\bibitem[{{Velovi{\'c}} {et~al.}(2023){Velovi{\'c}}, {Cotton}, {Filipovi{\'c}}, {Norris}, {Barnes}, \& {Condon}}]{Velovic2023}
{Velovi{\'c}}, V., {Cotton}, W.~D., {Filipovi{\'c}}, M.~D., {et~al.} 2023, \mnras, 523, 1933

\bibitem[{{Wang} {et~al.}(2023){Wang}, {Reville}, {Mizuno}, {Rieger}, \& {Aharonian}}]{Wang2023}
{Wang}, J.-S., {Reville}, B., {Mizuno}, Y., {Rieger}, F.~M., \& {Aharonian}, F.~A. 2023, \mnras, 519, 1872

\bibitem[{{We{\.z}gowiec} {et~al.}(2024){We{\.z}gowiec}, {Jamrozy}, {Chy{\.z}y}, {Hardcastle}, {Ku{\'z}micz}, {Heald}, \& {Shimwell}}]{Wezgoweiec2024}
{We{\.z}gowiec}, M., {Jamrozy}, M., {Chy{\.z}y}, K.~T., {et~al.} 2024, \aap, 691, A193

\bibitem[{{Willis} \& {Strom}(1978)}]{Willis1978}
{Willis}, A.~G. \& {Strom}, R.~G. 1978, \aap, 62, 375

\bibitem[{{Winner} {et~al.}(2019){Winner}, {Pfrommer}, {Girichidis}, \& {Pakmor}}]{Winner2019}
{Winner}, G., {Pfrommer}, C., {Girichidis}, P., \& {Pakmor}, R. 2019, \mnras, 488, 2235

\bibitem[{{Young} {et~al.}(2024){Young}, {Turner}, {Shabala}, {Stewart}, \& {Yates-Jones}}]{Young2024}
{Young}, S.~A., {Turner}, R.~J., {Shabala}, S.~S., {Stewart}, G. S.~C., \& {Yates-Jones}, P.~M. 2024, arXiv e-prints, arXiv:2412.14433

\end{thebibliography}




\end{document}